\documentclass[]{aa}

\usepackage{graphicx}
\usepackage{subcaption}
\usepackage{mwe}

\usepackage{txfonts}

\usepackage{hyperref}
\usepackage{url}
\usepackage{xcolor}
\hypersetup{
colorlinks,
linkcolor={red!80!black},
citecolor={blue!80!black},
urlcolor={blue!80!black}
}

\def\cm3{cm$^{-3}$}
\def\kms{km~s$^{-1}$}
\def\msunyr{M$_{\odot}$\,yr$^{-1}$}
\def\lsun{L$_{\odot}$}

\def\msun{M$_{\odot}$}
\def\zsun{Z$_{\odot}$}

\def\one{\ts {\,\sc i}}
\def\two{\ts {\,\sc ii}}
\def\three{\ts {\,\sc iii}}

\def\beq{\begin{equation}}
\def\eeq{\end{equation}}

\def\lesssim{\mathrel{\hbox{\rlap{\hbox{\lower4pt\hbox{$\sim$}}}\hbox{$<$}}}}
\def\gtrsim{\mathrel{\hbox{\rlap{\hbox{\lower4pt\hbox{$\sim$}}}\hbox{$>$}}}}

\def\ha{H$\alpha$}
\def\one{{\,\sc i}}
\def\two{{\,\sc ii}}
\def\three{{\,\sc iii}}

\def\oidoub{[O\one]\,$\lambda\lambda$\,$6300,\,6364$}

\newcommand{\code}[1]{\texttt{#1}}

\newcommand{\cmfgen}{\code{CMFGEN}}

\def\ergs{erg\,s$^{-1}$}

\def\sn{SN\,1987A}

\newcommand{\iso}[2]{\ensuremath{^{#1}\rm{#2}}}

\def\apjs{ApJS}
\def\apjl{ApJL}
\def\aap{A\&A}

\def\nifs{\iso{56}Ni}

\def\lya{Ly\,$\alpha$}
\def\ha{H$\alpha$}

\def\oidoub{[O\one]\,$\lambda\lambda$\,$6300.3,\,6363.8$}

\def\naid{Na\one\,$\lambda\lambda\,5896,5890$}

\def\mgifs{Mg\one]\,$\lambda$\,$4571.1$}

\def\neiifs{[Ne\two]\,12.810\,$\mu$m}

\def\mej{$M_{\rm ej}$}
\def\ekin{$E_{\rm kin}$}
\def\minit{$M_{\rm init}$}
\def\mic{\,$\mu$m}

\begin{document}

\title{A panchromatic photometric and spectroscopic modeling of SN\,1987A from 200 to 800\,d}
\titlerunning{Radiative transfer modeling of SN\,1987A}

\author{
Luc Dessart \inst{\ref{inst1},\ref{inst1a}}
}

\institute{
    French-Chilean Laboratory for Astronomy, IRL 3386, CNRS and Instituto de Astrofísica,
       Pontificia Universidad Católica de Chile, Casilla 306, Santiago, Chile.\label{inst1}
\and
    Institut d'Astrophysique de Paris, CNRS-Sorbonne Universit\'e, 98 bis boulevard Arago, F-75014 Paris, France\label{inst1a}
}

   \date{Received; accepted}

  \abstract{
Type II supernovae (SNe) at 1--3\,yr after explosion can document not only the preSN mass loss history of massive stars but also the chemistry of the inner ejecta rich in metals and eventually molecules and dust. With the aim of modeling \sn, we present a grid of radiative-transfer models of Type II SNe at \zsun/3, covering an initial mass range between 11 and 21\,\msun, a fixed ejecta kinetic energy of $1.2\times 10^{51}$\,erg, and $\sim$\,0.08\,\msun\ of \nifs. The influence of metallicity at such late times is found to be relatively weak. In this model set, the \oidoub\ flux at 1--2\,yr varies non-monotonically with \minit. It first rises with \minit\ due to the growing O abundance but drops for $M>17-19$\,\msun\ due to the higher ejecta density, with little change to the rest of the optical spectrum. Assuming CO molecules account for 90\,\% of the cooling in the C/O shell leads to a 10-20\,\% reduction in the [O\one] flux, equivalent to an O(1\,\msun) reduction in \minit\ at $M<19$\,\msun. The fractional [O\one] flux rises from 200 to $\sim$\,600\,d but decreases beyond as more decay power is channeled in, for example, the H-rich gas, and alternate coolants kick in. We obtain a good agreement to \sn\ for all \minit\ apart from [O\one], which appears in the optical as the only discriminant between models. The 13\,\msun\ model yields the best match, from UV to IR, but a higher \minit\ may result with allowance for strong  CO and SiO cooling. Past 500\,d, a growing amount of dust below $\sim$\,2000\,\kms\ is essential to account for the $\sim$\,1.5\,mag optical excess fading relative to predictions from decay power. Adopting 0.1\,\mic\ silicate grains, we find a dust mass of $10^{-4}$\,\msun\ at 600\,d rising to a few $10^{-3}$\,\msun\ at 700--800\,d -- this ejecta dust is optically thick from UV to IR. We also found that the 200--800\,d International Ultraviolet Explorer spectra of \sn\ are all contaminated by a background hot star, which also causes the relative UV brightening past 500\,d. Such contamination may complicate the interpretation of more distant SNe II at $\gtrsim$\,2\,yr once or if they fade to O($10^5$\,\lsun).
}

   \maketitle

\section{Introduction}

Recent years have seen a renewed interest in the late-time evolution of Type II supernovae (SNe) past the nominal age of one year after explosion. Rather than continuously fading according to the declining decay power, a number of events have revealed a knee in their optical light curve, with a flattening occurring typically at about 600\,d. This photometric change is associated with a dramatic change in spectral morphology from relatively narrow lines forming in the inner metal-rich ejecta to broad, boxy profiles forming in the outer ejecta (e.g., SN\,2004et, \citealt{kotak_04et_09}; SN\,2013ej, \citealt{mauerhan_13ej_17}; SN\,2017eaw, \citealt{weil_17eaw_20}). This transformation is thought to result from a shift in the dominant powering mechanism, that is from decay power early on to interaction power later on \citep{dessart_late_23}. The two nearby SNe 2023ixf and 2024ggi have documented this phenomenon and its diversity further and better than ever before. SN\,2023ixf exhibited signatures of interaction throughout the photospheric and nebular phase, both in the optical and in the ultraviolet (UV), which persist until today (e.g., \citealt{bostroem_23ixf_24}; \citealt{folatelli_23ixf_25}; \citealt{kumar_23ixf_25}; \citealt{bostroem_uv_26}; \citealt{dessart_23ixf_phot_26}). SN\,2024ggi showed signs of interaction with optically-thick circumstellar material early on \citep{wynn_24ggi_24,shrestha_24ggi_24,zhang_24ggi_24} but little sign of interaction during the photospheric phase and the nebular phase prior to about 500\,d  \citep{baron_24ggi_25,ferrari_24ggi_25,dessart_24ggi_25}. Only later at $\sim$\,600\,d were clear signatures of interaction detected in SN\,2024ggi \citep{hueichapan_24ggi_26}. All such cases reveal the complicated mass-loss histories of Type II SN progenitors, with mass-loss rates well above expectations for red-supergiant stars (e.g., 10$^{-4}$\,\msunyr\ for SN\,2023ixf; \citealt{nayana_23ixf_25}; \citealt{wynn_sed_25}; \citealt{bostroem_uv_26}).

While this new window on late-stage massive star mass loss is of much interest, there is also a greater need to better understand the emission from the inner ejecta powered by radioactive decay alone. Such signatures can help constrain the progenitor mass and thus document the connection between mass loss and initial mass \citep{fransson_chevalier_89,jerkstrand_04et_12,dessart_sn2p_21}. The optical spectrum at intermediate times of 300--600\,d is also a mix of two emitting regions (i.e., inner and outer ejecta; \citealt{dessart_late_23}; \citealt{dessart_23ixf_neb_26}) and thus a proper modeling of each contribution is needed to understand the whole. Such modeling can also shed light on the formation, in unshocked metal-rich gas, of molecules \citep{catchpole_87A_88,spyromilio_co_87A_88,bouchet_87A_89,aitken_mir_87A_88,roche_sio_87A_91,liu_dalgarno_92} and dust \citep{lucy_87A_91,wooden_87A_ir_93,meikle_87A_93}, as observed in \sn. A final interest is to extend previous radiative-transfer simulations of nebular-phase SNe II to metallicities below the solar value -- we start here with simulations at \zsun/3.

\sn\ remains undoubtedly the most studied SN to date and we will only briefly summarize the key relevant information for the work presented here on its nebular phase evolution (for a comprehensive review on \sn, see, for example, \citealt{sn1987A_rev_90}). \sn\ was monitored photometrically out to a few years \citep{hamuy_87A_88,hamuy_87A_90,bouchet_87A_91,walker_suntzeff_87A_91,bouchet_danziger_93}, from which optical, infrared (IR), and tentative bolometric light curves could be constructed (\citealt{suntzeff_87A_92}; this led to the identification of \iso{57}Co as an extra power source past $\sim$\,1000\,d). Spectroscopic monitoring was carried out at high cadence for several years and covering not just the optical, but also the near- and mid-IR (\citealt{catchpole_87A_87,catchpole_87A_88}; \citealt{phillips_87A_90}; \citealt{spyromilio_co_87A_88}; \citealt{roche_sio_87A_91}; \citealt{spyromilio_87A_91}; \citealt{meikle_87A_93}), as well as the UV with the International Ultraviolet Explorer (IUE; \citealt{pun_iue_87A_95}). These data revealed the formation of CO and SiO molecules after about 200\,d, but also of dust through the identification of emission profile asymmetry and optical fading after about 500\,d \citep{lucy_dust_89}.

Most of the modeling of the SN radiation has focused on the light curve, from which constraints on the ejecta mass, \nifs\ mass, as well as the \nifs\ mixing have been made (see, for example, \citealt{woosley_87A_late_88}; \citealt{blinnikov_87A_00}; \citealt{eastman_87A_89}; \citealt{utrobin_87A_93}). Early on, fluid instabilities developing after envelope shock passage have been recognized as a key ingredient for chemical mixing \citep{muller_87A_91,ebisuzaki_87A_89}, augmented by the intrinsic asymmetries of the neutrino-driven explosion mechanism (see, for example, \citealt{wongwathanarat_15_3d}), in conformity, for example, with spatially resolved observations of \sn\ taken many years after explosion \citep{abellan_87A_17}. The merger scenario is currently favored for \sn\ (e.g., \citealt{menon_heger_87A_17}).

The modeling of line emission and gas properties in \sn\ at nebular times allowed to set some constrains on the progenitor mass (i.e., 15--25\,\msun), the physics of these ejecta at late times, as well as the inhomogenous, clumpy structure of these metal-rich inner ejecta \citep{fransson_neb_87A_87,li_87A_93,KF98a,KF98b}. Modeling of the CO first overtone and fundamental constrained the amount of CO (see, e.g., \citealt{spyromilio_co_87A_88,liu_dalgarno_92}). Modeling of the emission line profiles past 500\,d constrained the location and amount of newly-formed dust in the inner ejecta (see, e.g., \citealt{lucy_dust_89}; \citealt{bevan_barlow_16}; \citealt{dessart_dust_25}).

Here, we revisit previous work on the nebular-phase properties of \sn\ by performing detailed radiative-transfer calculations, including dust at times, to model the UV to IR properties of \sn\ between 200 and 800\,d after explosion. In the next section, we present the numerical setup for our work, including the stellar evolution models performed at \zsun/3, the treatment of the explosion, and the radiative transfer for the resulting ejecta. In Section~\ref{sect_res}, we present the ejecta, photometric, and spectral properties for our radiative-transfer models based on the explosion of stars born with a mass between 11 and 21\,\msun. This includes a discussion of the nonmonotonicity of the \oidoub\ flux with \minit\ (Section~\ref{sect_oi}), an assessment of the impact of CO cooling (Section~\ref{sect_co}), as well as the impact of metallicity (Section~\ref{sect_z}). We then compare in Section~\ref{sect_87A} our results to the observations of \sn\ between 200 and 800\,d after explosion, considering the ``raw'' models as well as counterparts in which some 0.1\,\mic\ silicate dust is introduced in the inner ejecta below 2000\,\kms.
We conclude in Section~\ref{sect_conc}. Additional illustrations are presented in Appendix~\ref{sect_appendix}.

\begin{table*}
  \caption{Summary of ejecta properties used as initial conditions for our radiative-transfer calculations. In addition to these models at $Z_\odot/3$, we include the model s15.2 at $Z_\odot$ \citep{sukhbold_ccsn_16,dessart_sn2p_21} for comparison to th15 (see Section~\ref{sect_setup} for discussion.)
\label{tab_cmfgen_init}
}
\begin{center}
\begin{tabular}{
l@{\hspace{1mm}}c@{\hspace{1mm}}c@{\hspace{1mm}}c@{\hspace{1mm}}c@{\hspace{1mm}}c@{\hspace{1mm}}c@{\hspace{1mm}}c@{\hspace{1mm}}c@{\hspace{1mm}}c@{\hspace{1mm}}c@{\hspace{1mm}}c@{\hspace{1mm}}c@{\hspace{1mm}}c@{\hspace{1mm}}c@{\hspace{1mm}}c@{\hspace{1mm}}c@{\hspace{1mm}}
}
\hline
Model   &  $E_{\rm kin}$ & $M_{\rm ej}$   & He$_{\rm c}$ & H  & He  & \iso{12}C &  \iso{16}O & Ne & \iso{24}Mg & \iso{28}Si&  \iso{40}Ca & \iso{52}Fe&  \nifs& \iso{57}Ni &  \iso{58}Ni \\
&  [foe]     & [\msun] & [\msun] & [\msun] & [\msun] & [\msun] & [\msun] & [\msun] & [\msun] & [\msun] & [\msun] & [\msun] & [\msun] & [\msun] & [\msun] \\
\hline
s15.2   &  0.84  &    10.95   &    4.46 &  5.24 &  3.93  & 1.70(-1)  & 1.00  & 2.20(-1)  & 4.41(-2)  & 8.65(-2)  & 7.77(-3)  & 6.95(-4)  & 6.33(-2)  &   2.79(-3)  &  3.68(-3) \\
 th15   &  1.18  &    12.73   &    4.73 &  6.30 &  4.35  & 1.82(-1)  & 1.18  & 2.59(-1)  & 4.77(-2)  & 1.48(-1)  & 1.67(-2)  & 1.20(-3)  & 7.70(-2)  &   2.33(-3)  &  1.83(-3) \\
\hline
 th11   &  1.17  &     9.13   &    3.27 &  5.21 &  3.22  & 9.63(-2)  & 0.32  & 5.08(-2)  & 2.49(-2)  & 5.80(-2)  & 4.32(-3)  & 8.54(-4)  & 7.84(-2)  &   2.71(-3)  &  1.48(-3) \\
 th13   &  1.17  &    10.82   &    4.04 &  5.67 &  3.75  & 1.51(-1)  & 0.76  & 1.41(-1)  & 3.67(-2)  & 1.10(-1)  & 9.24(-3)  & 1.19(-3)  & 7.79(-2)  &   2.31(-3)  &  2.61(-3) \\
 th15   &  1.18  &    12.73   &    4.73 &  6.30 &  4.35  & 1.82(-1)  & 1.18  & 2.59(-1)  & 4.77(-2)  & 1.48(-1)  & 1.67(-2)  & 1.20(-3)  & 7.70(-2)  &   2.33(-3)  &  1.83(-3) \\
 th17   &  1.17  &    14.23   &    5.54 &  6.71 &  4.89  & 1.92(-1)  & 1.67  & 3.67(-1)  & 6.96(-2)  & 1.13(-1)  & 1.02(-2)  & 1.37(-3)  & 7.61(-2)  &   2.00(-3)  &  1.53(-3) \\
 th19   &  1.19  &    15.30   &    6.29 &  7.16 &  5.46  & 3.02(-1)  & 1.67  & 2.08(-1)  & 4.34(-2)  & 2.06(-1)  & 1.09(-2)  & 2.01(-3)  & 7.67(-2)  &   1.42(-3)  &  1.16(-3) \\
 th21   &  1.21  &    17.10   &    7.14 &  7.35 &  5.96  & 2.51(-1)  & 2.58  & 1.42(-1)  & 4.83(-2)  & 3.73(-1)  & 1.53(-2)  & 1.74(-3)  & 7.86(-2)  &   1.76(-3)  &  1.02(-3) \\
\hline
\end{tabular}
\end{center}
{\bf Notes:} The columns give in turn the modelname (``s'' stands for solar metallicity, ``th'' stands for third solar metallicity, followed by the model initial mass in \msun), the kinetic energy, the ejecta mass, the progenitor He-core mass, and the ejecta yields of important specific isotopes in order of increasing atomic mass (i.e., H, He, \iso{12}C, \iso{16}O, Ne, \iso{24}Mg, \iso{40}Ca, \iso{48}Cr, \iso{52}Fe, \nifs, \iso{57}Ni, and \iso{58}Ni). For He, we give the sum of \iso{3}He and \iso{4}He, and for Ne we give the sum of \iso{20}Ne and \iso{22}Ne. Also, 1\,foe\,$\equiv$\,10$^{51}$\,erg.
\label{table_init}
\end{table*}

\section{Numerical setup}
\label{sect_setup}

\subsection{PreSN evolution and explosion}

The massive-star evolution models presented in this work were performed in a similar fashion to that described in \citet{sukhbold_ccsn_16}, with a few specific changes. All simulations were run with solar abundances scaled by a factor of $1/3$, that is all elements from carbon and above have their abundance multiplied by $1/3$. Rotation was ignored. The parameter $\alpha_{\rm MLT}$ controlling convection was set to two. The wind mass loss rates were adopted from the recipes given in \citet{nieuwenhuijzen_mdot_90}, with an additional metallicity scaling of $(Z/Z_\odot)^{0.5}$ or 0.577 for the present case.

The nuclear network included species up to bismuth ($Z=$\,83) and carried a maximum of 1703 isotopes in roughly 2000 zones. The detailed composition was evolved in a coprocessing mode. That is, the abundances were updated each zone cycle for the existing temperature, density, and convection, although the energy generation from the big nuclear network was not directly coupled to the stellar structure calculation. Energy generation for the stellar structure came from the 19 isotope network up until oxygen depletion and from the silicon quasi-equilibrium network after that. At the time of core collapse, all these models were red supergiant stars. While this would be unsuitable for the modeling of the photospheric phase properties of a Type II-peculiar SN like 1987A, it is of secondary importance for the nebular-phase modeling presented here (see, e.g., \citealt{dessart_sn2p_21}).

At the time of core collapse, the stars were exploded using a piston placed at the Lagrangian mass where the total entropy moving out first exceeded 4.0\,k$_{\rm B}$ per baryon. The piston was moved to impart the desired kinetic energy near $1.2 \times 10^{51}$\,erg. Because these models produced a \nifs\ mass of 0.1\,\msun\ or more, the inner ejecta layers were excised such that the ejecta used for the radiative transfer calculations would all have 0.07--0.08\,\msun\ as suitable for \sn\ (see, e.g., \citealt{woosley_87A_late_88}). The full grid contains explosion models from initial stellar masses between 11 and 21\,\msun, ejecta masses between 9.1 and 17.1\,\msun, a \nifs\ mass of $\sim$\,0.075\,\msun, and an O yield between 0.3 and 2.6\,\msun\ -- a summary of this model set is presented in Table~\ref{tab_cmfgen_init}.

Compared to the models of \citet{sukhbold_ccsn_16}, some differences arise (see th15 vs. s15.2 in the top rows of Table~\ref{tab_cmfgen_init}). Some of these are well understood. For example, the lower metallicity leads systematically to smaller mass loss rates and a larger preSN mass than the solar metallicity models computed in \citet{sukhbold_ccsn_16}. This work also used updated pair neutrino losses \citep{sukhbold_evol_18} that differ from those employed in \citet{sukhbold_ccsn_16}. However, any small change, for example in zoning or time step, can send the star down a different evolutionary path \citep{farmer_mesa_16,sukhbold_evol_18}, such that two stars with similar initial masses can have different preSN structures and fate. Hence one should not expect models s15.2 and th15 to have identical properties at core collapse.

\subsection{Radiative transfer}

Using the ejecta models described in the previous section as initial conditions, we performed 1D nonlocal thermodynamic equilibrium time-dependent radiative transfer from 200 to 800\,d with \cmfgen\ \citep{HD12}. In order to allow for macroscopic mixing in spherical symmetry (and without microscopic mixing) the original unmixed composition structure at 200\,d was transformed into a shuffled-shell structure following the procedure described in \citet{dessart_shuffle_20}. This complex structure was sampled with a total of about 350 grid points to finely resolve all composition gradients as well as the optical-depth scale. The \nifs-bubble effect (e.g., \citealt{woosley_87A_late_88}), which leads to a strong expansion of \nifs-rich regions but a modest compression of the surrounding material (typically a factor of two), was ignored (see discussion in Section~5 of \citealt{dessart_sn2p_21}). Interaction with CSM was ignored such that the only power source at those nebular times was radioactive decay. Here, we accounted for the five two-step decay chains whose leading isotope are (in order of increasing atomic mass) \iso{44}Ti, \iso{48}Cr, \iso{52}Fe, \nifs, and \iso{57}Ni. In order of importance, the \nifs\ decay chain dominates at all times considered, but the \iso{57}Ni-decay chain is no longer negligible after about 500\,d (and would dominate after about 1500\,d; see Fig.~1 of \citealt{dessart_sn2p_21}). Nonthermal processes were treated as by \citet{li_etal_12_nonte} and \citet{d12_snibc}. The nonlocal energy deposition was calculated by solving the radiative-transfer equation with a gray absorption-only opacity to $\gamma$-rays set to 0.06 $Y_{\rm e}$\,cm$^2$\,g$^{-1}$. Ejecta clumping was ignored in this work.

Since the ejecta gas conditions are cool and at most partially ionized, we limited the model atom to ionization states {\sc i}-{\sc ii}-{\sc iii} for most species. In order of increasing atomic weight and then ionization, we included H\one\ (26,36), He\one\ (40,51), He\two\ (13,30), C\one\ (14,26), C\two\ (14,26), N\one\ (44,104), N\two\ (23,41), O\one\ (21,51), O\two\ (54,123), Ne\one\ (78,155), Ne\two\ (22,91), Ne\three\ (32,80), Na\one\ (22,71), Mg\one\ (39,122), Mg\two\ (31,80), Si\one\ (100,187), Si\two\ (31,59), S\one\ (106,322), S\two\ (56,324), Ar\one\ (56,110), Ar\two\ (134,415), K\one\ (25,44), Ca\one\ (76,98), Ca\two\ (21,77), Sc\one\ (26,72), Sc\two\ (38,85), Sc\three\ (25,45), Ti\two\ (37,152), Ti\three\ (33,206), Cr\two\ (28,196), Cr\three\ (30,145), Fe\one\ (413,1142), Fe\two\ (228,2698), Fe\three\ (96,1001), Co\two\ (112,1005), Co\three\ (88,1075), Ni\one\ (56,301), Ni\two\ (59,1000), Ni\three\ (47,1000), Ba\one\ (17,31), and Ba\two\ (28,61). Here, the numbers in parentheses correspond to the number of super levels and full levels (for additional information on super levels, see \citealt{hm98}).

Molecules are ignored in this work. In Section~\ref{sect_co}, we explore the potential impact of CO molecule formation on our results, in particular on the strength of the \oidoub\ line flux. Dust radiative transfer is however performed since it is a critical driver of the evolving radiative properties of \sn\ after about 500\,d \citep{lucy_dust_89}. Practically, we used the same formalism as in \citet{dessart_dust_25}. Dust is introduced as a post-processing step in a given converged \cmfgen\ calculation by specifying its spatial distribution, type, and temperature. Exploratory calculations were presented in \citet{dessart_dust_25}, which are now extended to more epochs, but using the same spatical distribution within the inner 2000\,\kms\ of the ejecta, thus the metal-rich regions only. Theoretically, silicate grains are expected to form first and as small grains \citep{sarangi_dust_22} so we used 0.1\,\mic\ silicates. The opacities for these dust grains are adopted from \citet{sarangi_dust_22}, and originally from \citet{draine_li_07}, \citet{weingartner_draine_01} and \citet{zubko_dust_04}. In this work, we constrained the dust present in the ejecta  primarily through its combined absorptive and scattering influence on the spectral energy distribution, thus considering the impact of dust on all line profiles between the UV and IR. Given the many degeneracies in this exercise, we did not explore with alternate dust types -- we tried with a larger grain size of 1.0\mic\ but this produced little difference.

\begin{figure}
\centering
\includegraphics[width=\hsize]{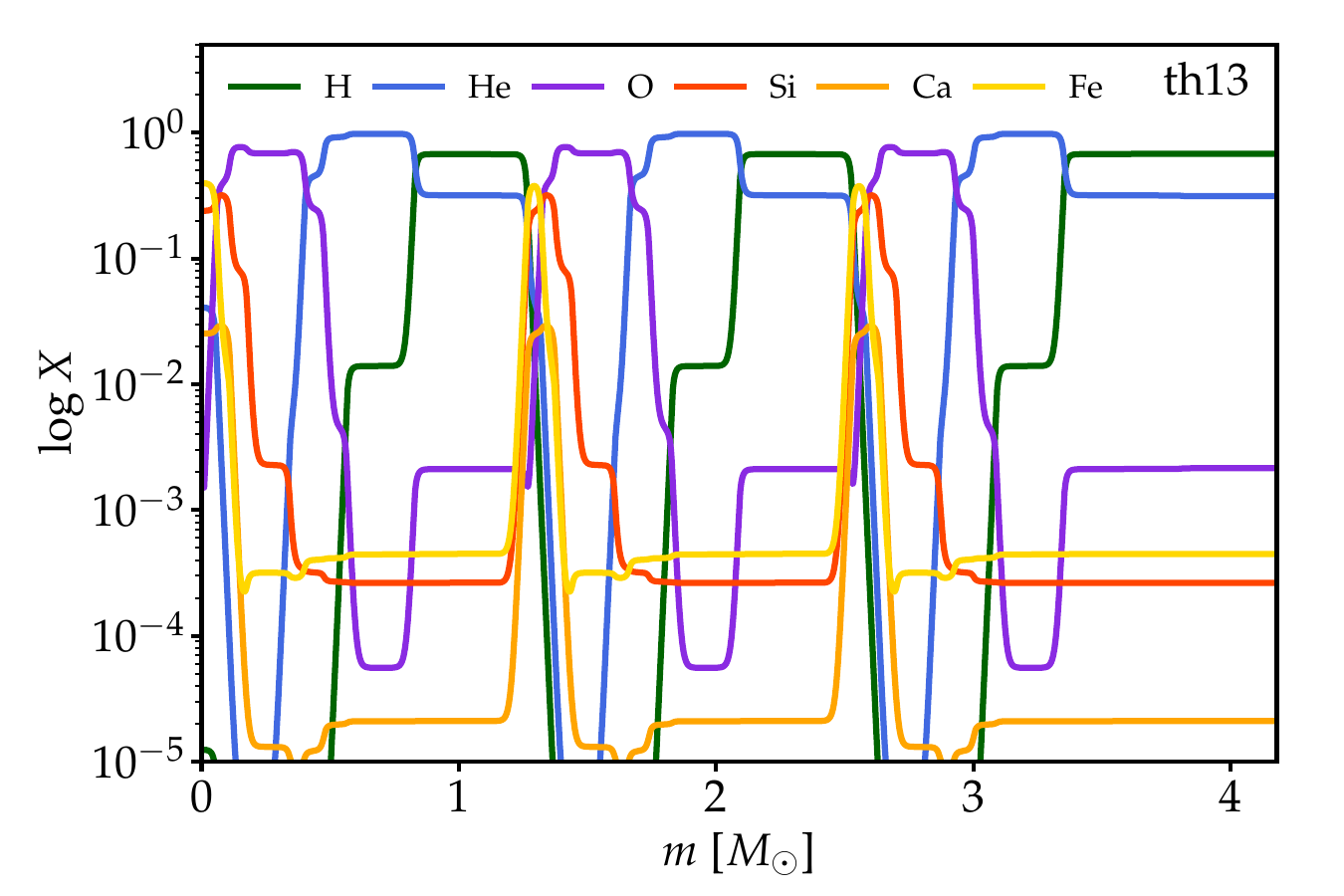}
\caption{Shuffled-shell composition profile versus Lagrangian mass for H, He, O, Si, Ca, and Fe in model th13 at 200\,d and used as initial conditions for the radiative transfer calculations.}
\label{fig_prog}
\end{figure}

\newpage

\section{Ejecta, photometric, and spectral properties for our model set th11 to th21}
\label{sect_res}

A detailed presentation of nebular-phase phase properties of Type II SN gas and radiation properties was provided in \citet{dessart_sn2p_21} and \citet{dessart_ir_25} -- see also \citet{fransson_chevalier_89} and \citet{jerkstrand_04et_12}. In this section, we focus only on a few specific items and present additional results alongside the comparison to \sn\ in Section~\ref{sect_87A}. We used model th13 for most of the present discussion since it best matches \sn.

\subsection{Ejecta properties}

\begin{figure}
\centering
\includegraphics[width=\hsize]{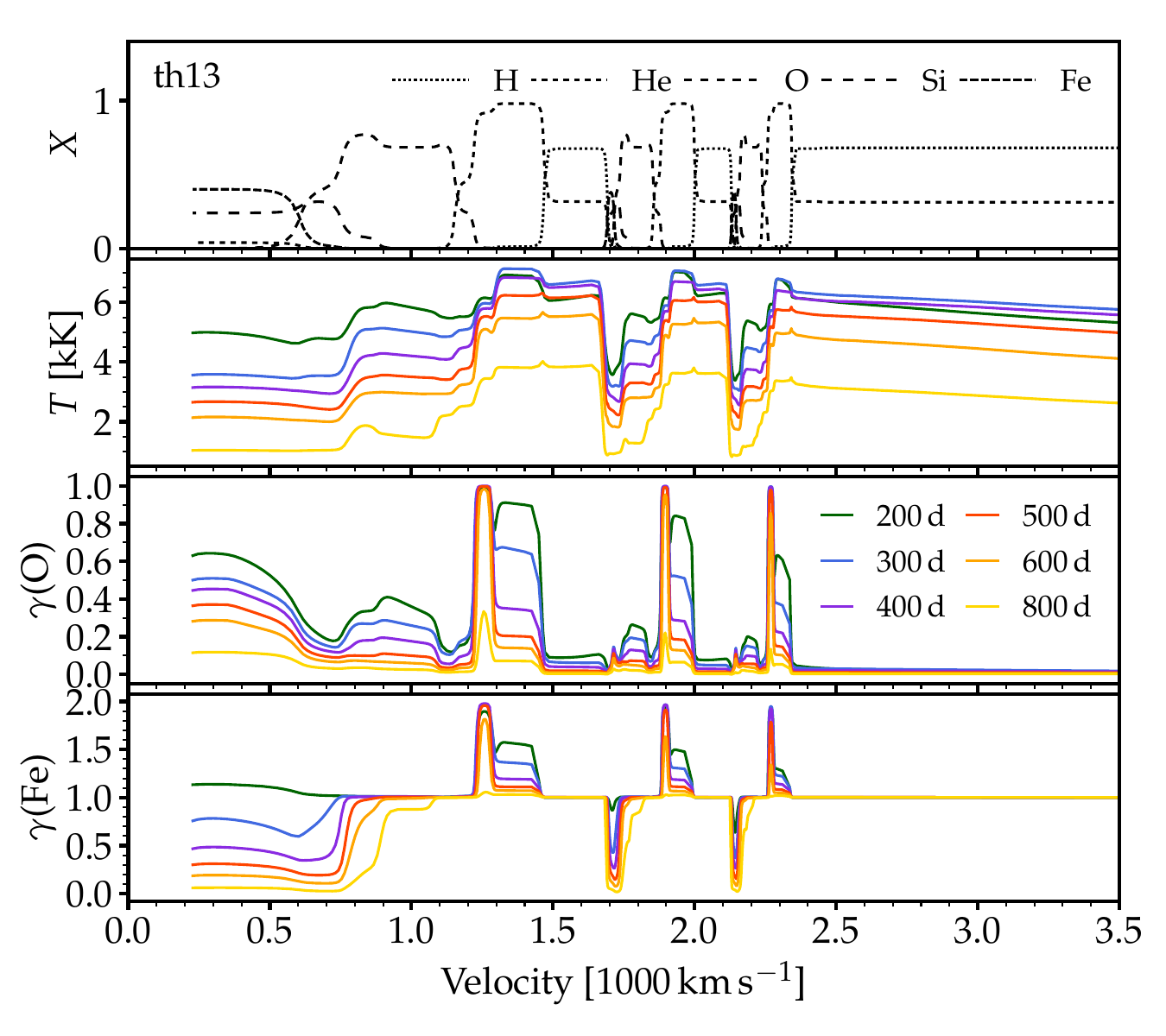}
\caption{Evolution of ejecta properties for model th13 from 200 to 800\,d. From top to bottom, we show the composition of the main species followed by the temperature and ionization of O and Fe. A reported ionization of zero corresponds to a neutral state, one to once ionized etc. We only show the inner ejecta where the bulk of the decay power is absorbed.
\label{fig_th13_ej}
}
\end{figure}

Figure~\ref{fig_th13_ej} shows the evolution of ejecta properties for model th13 from 200 to 800\,d. Conditions are marginally nebular at 200\,d (total ejecta electron-scattering optical depth $\tau_{\rm es}$ of 1.28) and eventually fully nebular at 800\,d (i.e., $\tau_{\rm es}=$\,0.011). The temperature is initially $\sim$\,5000\,K, with the He-rich and H-rich regions above 6000\,K and the Fe-rich regions below 4000\,K. This complicated temperature profile arises from the fact that the conditions are optically thin and the coolants vary from shell to shell -- more metal-rich regions typically have greater cooling power and thus have lower temperatures (Fig.~\ref{fig_cool_th13_350d_700d} in the appendix shows the dominant coolants in model th13 at 350 and 700\,d). As time passes, the temperature exhibits qualitatively the same profile but drops everywhere as the heating source (i.e., radioactive decay) exponentially declines. By the end of the simulation at 800\,d, the temperature minima are at 800--1000\,K and located in the Fe-rich shells while the maxima are around 3500\,K in the He-rich and H-rich regions. The O-rich material is at about 1500\,K.

The bottom two panels of Fig.~\ref{fig_th13_ej} illustrate the ionization profile of O and Fe versus velocity. Higher ionizations for all species are encountered in the He/N shell. In the O-rich shell, the O ionization is initially about 0.4 but drops to below 0.1 at 800\,d. Fe is once ionized throughout most of the ejecta, apart from the He/N shell (Fe$^{2+}$) and the Fe-rich regions (mostly neutral ion). These changes are relatively modest over such a long time span and will lead to a weak evolution of spectral properties with lines mostly from neutral and once ionized species.

\begin{figure}
\centering
\includegraphics[width=\hsize]{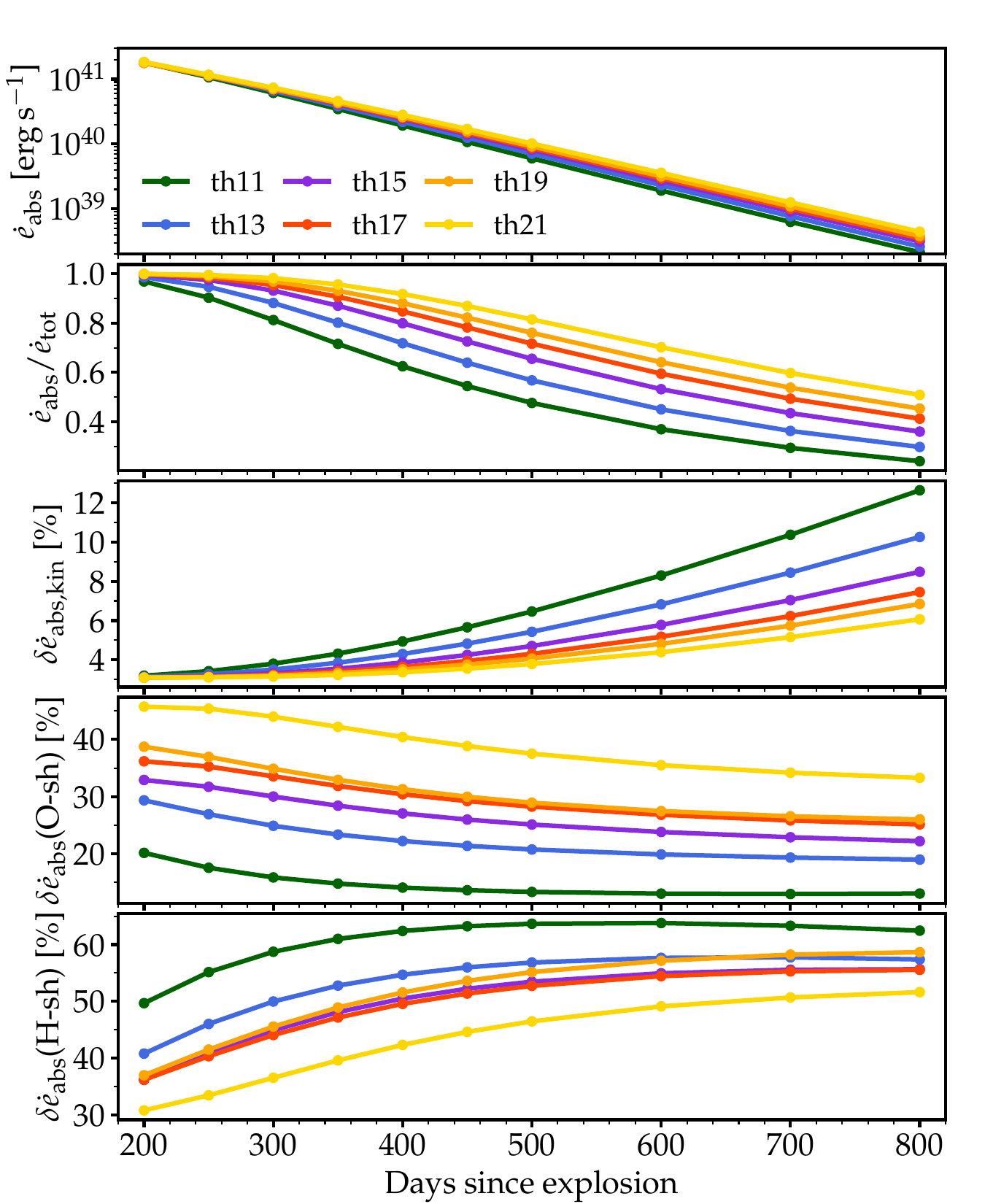}
\caption{From top to bottom, evolution of the total decay power absorbed by the ejecta, the fraction of the total power emitted that is absorbed, the percentage fraction of the decay power absorbed by the ejecta that is kinetic (i.e., positrons), and the percentage fraction of the decay power absorbed in the O-rich and H-rich shells.
\label{fig_edecay}
}
\end{figure}

Figure~\ref{fig_edecay} illustrates the evolution of the decay power absorbed in the ejecta for our set of models th11 to th21 (i.e., \minit\ of 11 to 21\,\msun). Although we included five two-step decay chains, the main contributor at those late times is \nifs, and to a lesser extent \iso{57}Ni (e.g., $\sim$\,4\,\% of the total at 800\,d in model th13). Full trapping of $\gamma$-rays holds at 200\,d but their escape starts at about that time such that only 20\,\% (50\,\%) of $\gamma$-rays are absorbed by the ejecta at 800\,d in models th11 (th21). Since all ejecta models have the same \ekin, same \nifs\ mass and distribution, these differences arise from the different \mej\ (i.e., 9.13 up to 17.10\,\msun\ from th11 to th21). The larger is the $\gamma$-ray escape, the greater is the contribution from positrons, which varies from $\sim$\,12\,\% in model th11 to $\sim$\,5\,\% in model th21. Similarly, the growth of the $\gamma$-ray mean free path with time leads to a reduction of the decay power absorbed in the O-rich gas (only present in the metal-rich inner ejecta) relative to the H-rich gas (present both in the inner, metal-rich ejecta and in the outer ejecta beyond 2000\,\kms).

\begin{figure}
\centering
\includegraphics[width=\hsize]{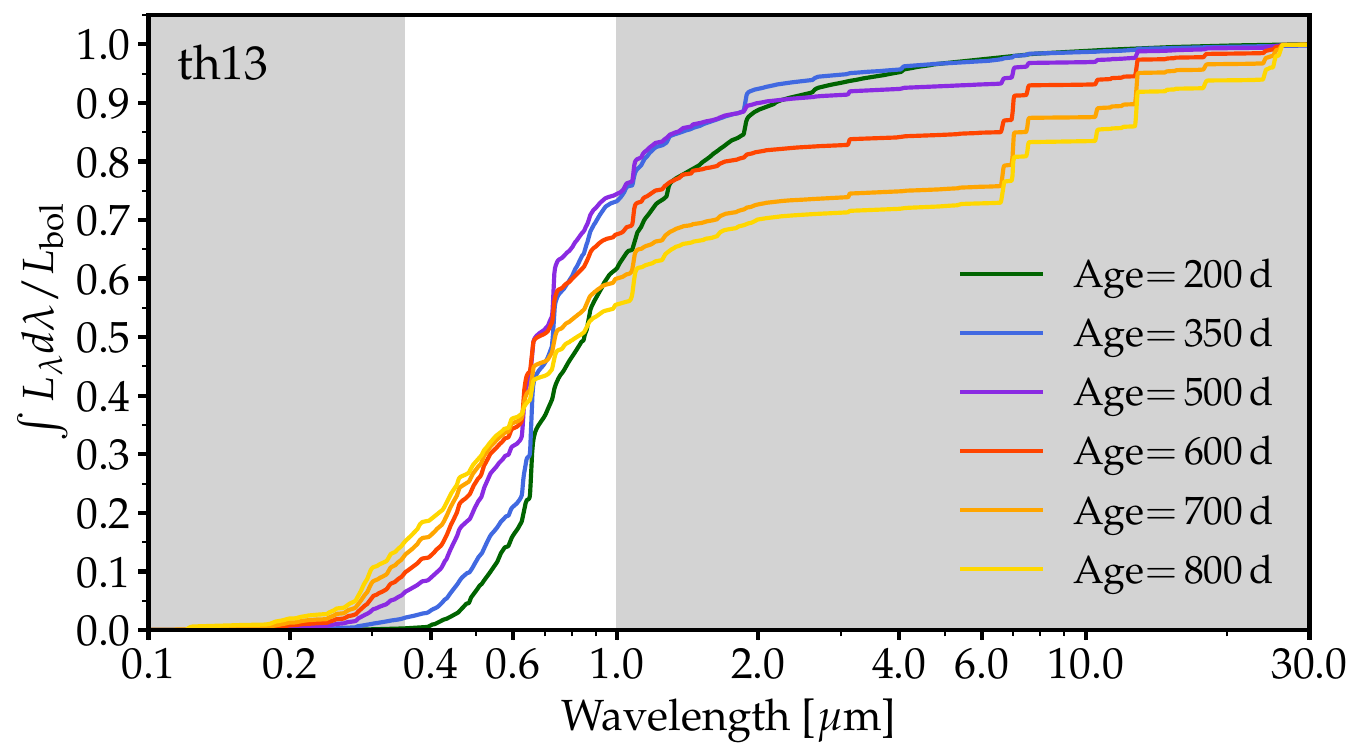}
\caption{Cumulative flux versus wavelength and time in the nebular phase for the model th13. A grey shading indicates the ultraviolet and IR ranges on each side of the optical range.
\label{fig_cum}
}
\end{figure}

These variations will lead to a strengthening of cooling from the H-rich gas over the 200 to 800\,d timespan, as well as a strengthening of emission from the Fe-rich gas where positrons are absorbed. Together with the changes in temperature and ionization, the spectral energy distribution arising from these models evolves significantly in time (Fig.~\ref{fig_cum}). In model th13, 62\,\%\ (38\,\%) of the emergent flux is channeled into the optical between 3500\,\AA\ and 1\,\mic\ (infrared, or $>$\,1\,\mic) at 200\,d, with essentially zero flux in the UV below 3500\,\AA\ (Fig.~\ref{fig_cum}). At 800\,d, the fraction falling in the optical has dropped to 40\,\%, with about the same fraction in the IR (45\,\%) but now 15\,\% in the UV.

\subsection{The \oidoub\ doublet}
\label{sect_oi}

Of particular interest is the evolution of the \oidoub\ line flux at nebular times since it is routinely used to evaluate the progenitor mass (e.g., \citealt{fransson_chevalier_89}; \citealt{jerkstrand_04et_12}; \citealt{dessart_sn2p_21}). In the context of \sn, the analysis of the \oidoub\ line flux was also used to set constraints on the oxygen abundance, clumping, and optical depth effects (see, e.g., \citealt{spyromilio_pinto_91,li_mccray_o1_92,chugai_87A_94}).

In our simulations, following the increase in O abundance, or more specifically the size of the O-rich shell (i.e., the combination of the O/Si, O/Ne/Mg, and O/C shells), a greater fraction of the decay power is absorbed by the O-rich material for increasing \minit\ (Fig.~\ref{fig_edecay}). If the ionization is relatively low in the O-rich shell, \oidoub\ may be the dominant coolant. If the conditions are identical in the O-rich shell in ejecta arising from the explosion of stars of different \minit, that [O\one] flux will scale with \minit. However, complications may enter when the ionization is too high, in which case Fe\two\ may be an important coolant of the O-rich material, or too low, in which case \naid\ or \mgifs\ may take over -- this situation is however more relevant in stripped-envelope SNe \citep{dessart_snibc_21}. Another complication is related to the density. Forbidden lines thrive at low density, which is the reason behind the delayed emergence at about 200\,d of \oidoub\ in SNe II, and even more so in low-energy Type II SNe because of their slower expanding, denser inner ejecta (see, e.g., SN\,2005cs; \citealt{pastorello_etal_05cs2}).

\begin{figure}
\includegraphics[width=\hsize]{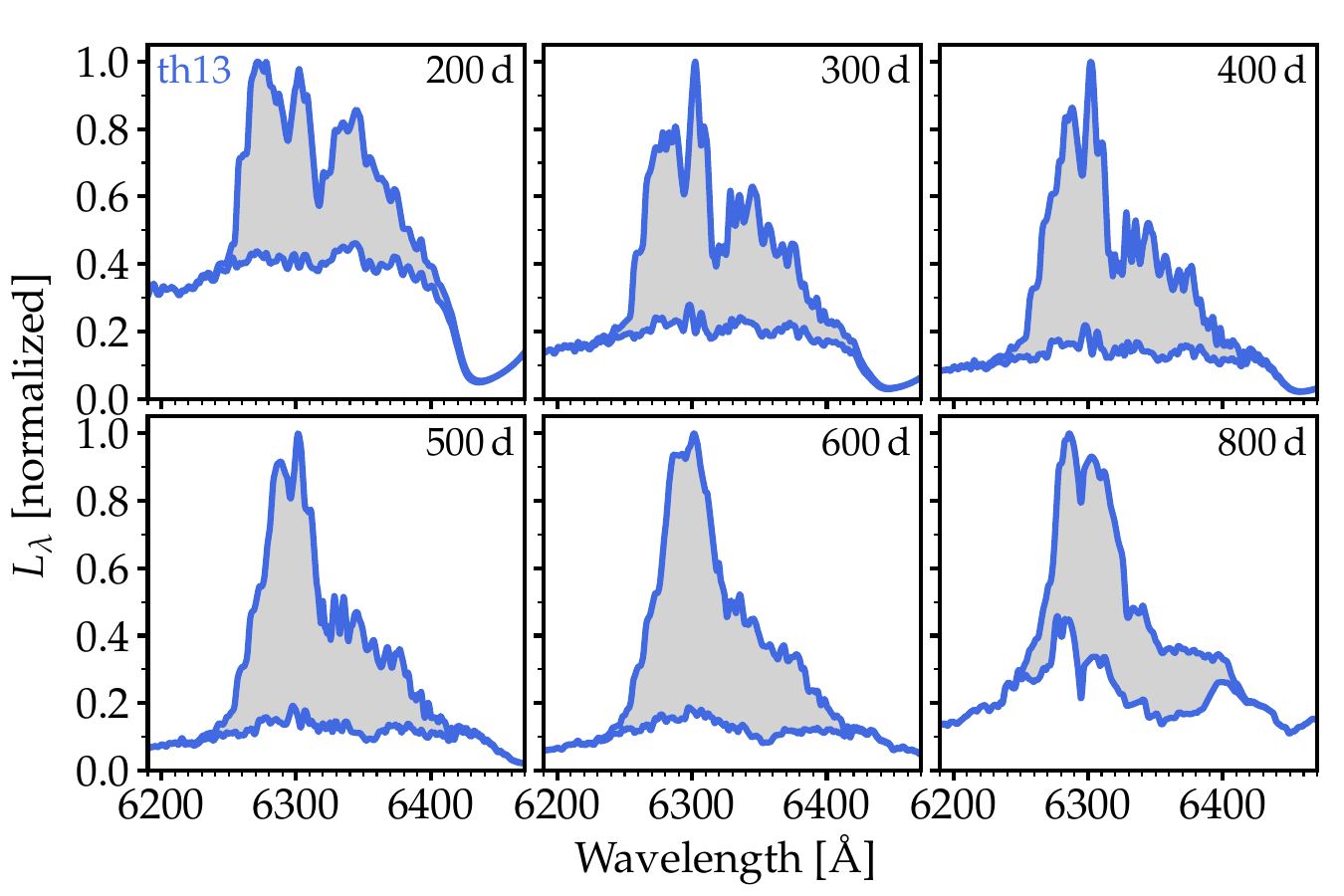}
\caption{Evolution of the \oidoub\ line flux (grey shading) for model th13 from 200 to 800\,d. The rest of the flux in this regions is due to Fe (primarily Fe\two), apart from the depression at $\gtrsim$\,4200\,\AA\ (and weakening with time passing) corresponding to the \ha\ trough. Time is ordered from left to right and top to bottom.
\label{fig_oi_th13}
}
\end{figure}

Furthermore, optical depth effects remain important even at those nebular epochs. First, $\tau_{\rm es}$ is still about unity in all models th11 to th21 at 200\,d, so electron scattering alone can introduce some effect. The lines themselves may remain optically thick out to late times. In model th13 at 200\,d, the line optical depth of [O\one]\,$\lambda$\,6300 is about 0.3 and thus barely optically thin whereas that of [O\one]\,$\lambda$\,6363 is about 0.1. The \oidoub\ doublet is also affected by overlapping Fe\two\ lines. As a result, the evolution of \oidoub\ from 200 to 800\,d is quite complicated and it is only at 500--600\,d that this emission appears strong, largely unaffected by Fe, and with the expected blue-to-red ratio of 3.1 (i.e., the ratio of the $A$ values for these transitions; Fig.~\ref{fig_oi_th13}). At times past 600\,d, the reduction in decay power absorbed in the O-rich gas boosts the Fe emission relative to \oidoub\ and alters its shape and flux significantly.

\begin{figure}
\includegraphics[width=\hsize]{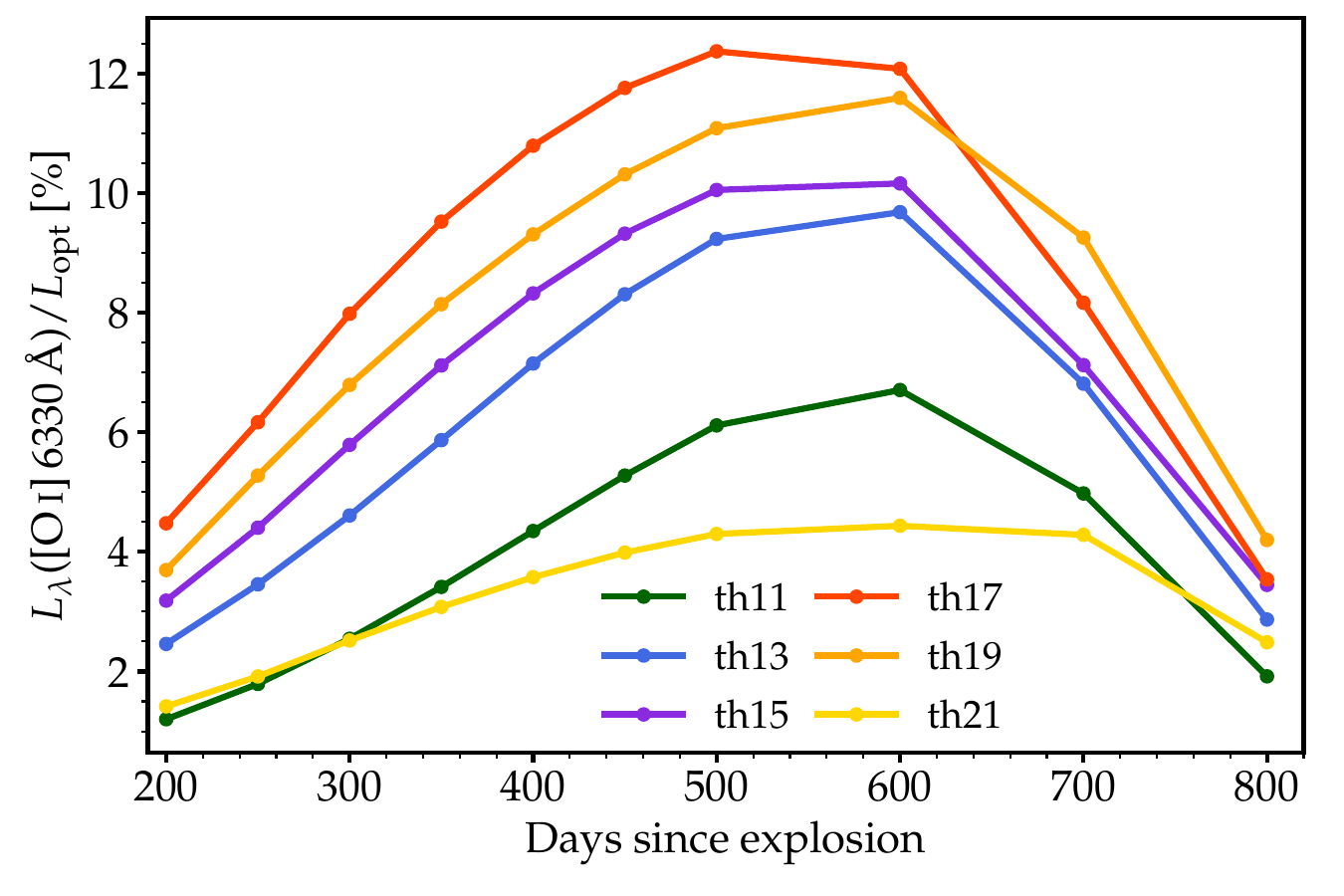}
\caption{Evolution of the fractional \oidoub\ line luminosity relative to total optical luminosity for our model set and given as a percentage. Filled dots indicate the epochs at which the models were computed.
\label{fig_oi_sum}
}
\end{figure}

Figure~\ref{fig_oi_sum} shows the evolution of the fractional \oidoub\ luminosity (relative to the optical luminosity). It rises in all models until 500-600\,d as the ionization and density decline but eventually drops beyond as other coolants kick in (e.g., the boost to the UV flux [Fig.~\ref{fig_cum}]; lines in the IR such as \neiifs; see also \citealt{dessart_ir_25}). Although the [O\one] line flux increases with \minit\ in the mass range 11 to 17\,\msun, further increase of \minit\ leads to its decrease. The [O\one] line flux is in fact comparable for models th11 and th21 which stand 10\,\msun\ apart on the main sequence, and correspond to a final O yield of 0.32\,\msun\ (th11) compared to 2.58\,\msun\ (th21). The origin of this non-monotonicity is that despite the growing O yield with \minit, the density of the O-rich material also grows and eventually quenches the formation of forbidden lines such \oidoub. This highlights an important degeneracy of Type II SN nebular-phase spectra in their use as progenitor mass diagnostics.

\begin{figure*}
\includegraphics[width=0.49\hsize]{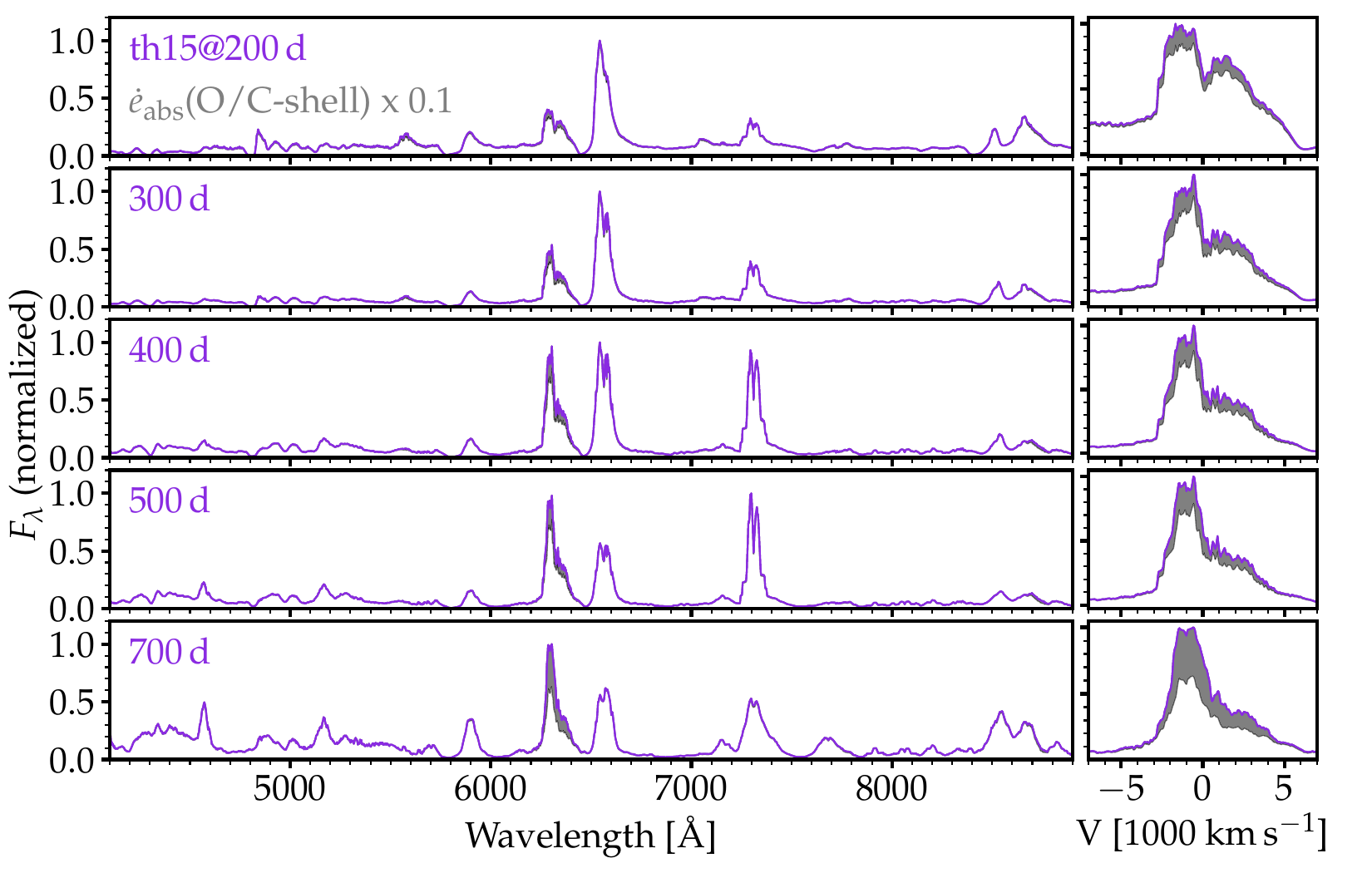}
\includegraphics[width=0.49\hsize]{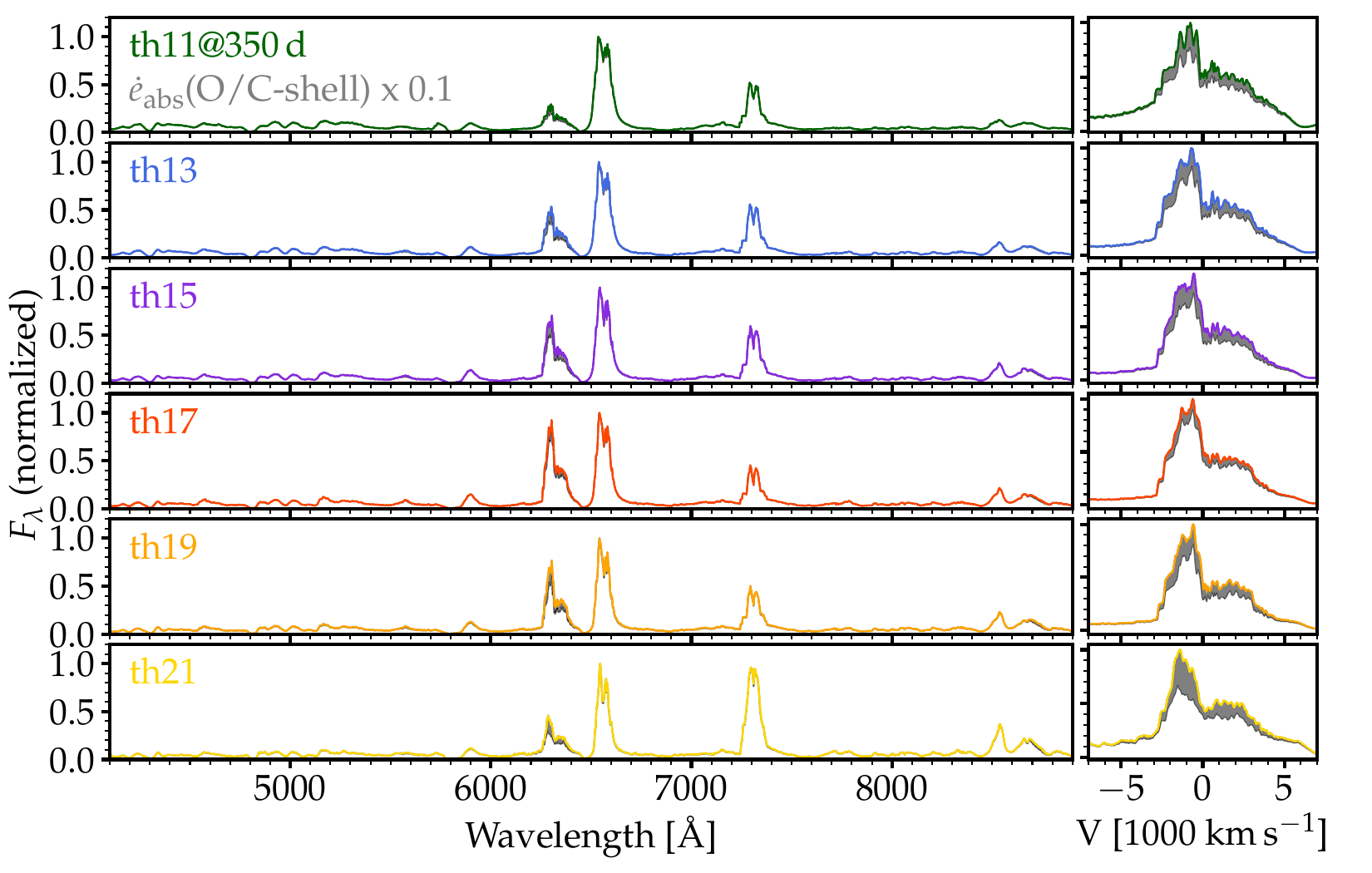}
\caption{Exploration of the potential impact of CO cooling on the model th15 from 200 to 700\,d (left) and on models th11 to th21 at 350\,d (right). In each subfigure, the left panel shows the optical spectrum and the right panel zooms in on the \oidoub\ region. CO cooling is mimicked here by assuming it accounts for a maximum of 90\,\% of the cooling of the C/O shell.
\label{fig_co_spec}
}
\end{figure*}

\subsection{Evaluating the influence of CO molecule formation and cooling}
\label{sect_co}

In this work as well as in most other works on nebular-phase modeling of SNe II, CO molecule formation and cooling are ignored in the global radiative transfer solution. \citet{liljegren_co_20} studied molecular formation (e.g., of CO and related molecules) in SNe but this calculation was not coupled to the full radiative transfer solution. Similarly, modeling of the CO first overtone and fundamental (see technique outlined in \citealt{mera_mol_26}) were performed by \citet{medler_22acko_26} for SN\,2022acko and by \citet{mera_24ggi_26} for SN\,2024ggi using spectra from  the James Webb Space Telescope. Exceptions to this are \citet{mcleod_co_24} for Type II SN models and \citet{liljegren_ibc_mol_23} for Type Ibc SN models. Furthermore, the impact of CO cooling is generally hard to gauge from SN observations since these typically capture the CO first overtone in the $K$ band (e.g., \citealt{rho_17eaw_18}) but generally lack the stronger fundamental at $\gtrsim$\,4.3\,\mic\ (e.g., \citealt{spyromilio_co_87A_88}), and even more rarely document the emission from SiO at longer wavelengths (see, for example, \citealt{roche_sio_87A_91}).

\begin{figure*}
\includegraphics[width=\hsize]{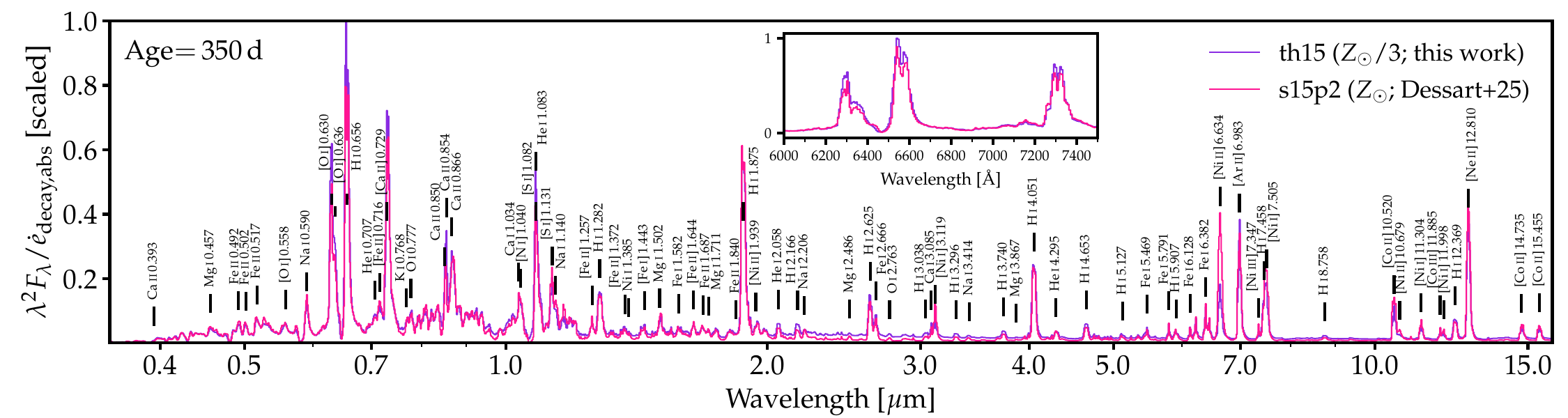}
\caption{Spectral comparison of the models th15 (at \zsun/3) and s15p2 (at \zsun) and at a post-explosion time of 350\,d. The spectra are first normalized to their associated decay-power absorbed at that time, scaled by $\lambda^2$ to yield a flat flux across optical and IR, and then divided by the maximum flux in model th15. The spectra were also rebinned to a resolution of 600\,\kms. Line identifications are provided for most emission features such that this figure can be used for line identifications elsewhere in the document.
\label{fig_th15_s15p2_350d}
}
\end{figure*}

But one can test the maximum impact of molecular cooling on SN spectra without doing any molecular modeling. Indeed,
in practice, CO cooling should ``radiate away'' a fraction of the decay power absorbed by the O-rich material and thus reduce the \oidoub\ line flux, thereby impacting the inferred progenitor mass. One can gauge the impact of CO cooling by assuming that CO is the main coolant of the O/C shell (the situation is more complicated since CO may be only a partial coolant, but of both the C/O shell and the O/Ne/Mg shell; \citealt{liljegren_co_20}). In \citet{dessart_24ggi_25}, we estimated the reduction on the \oidoub\ line flux based on the ratio $M$(O/C-shell)/$M$(O-shell), which was at most 30\,\% for $M>$\,12\,\msun\ in the massive star models of \citet{sukhbold_ccsn_16}.

Here, we directly evaluate the impact on the spectra by recalculating a sample of our models with a decay-power absorbed in the C/O shell scaled by a factor of 0.1 -- this is equivalent to assuming that CO is responsible for 90\,\% of the cooling in the C/O shell -- obviously this is not going to produce any CO emission in the IR but because it drains the power absorbed in the C/O shell it can mimic the impact CO would have if present. Figure~\ref{fig_co_spec} shows the corresponding optical spectra for model th15 from 200 to 700\,d (left column) and for models th11 to th21 at 350\,d. Surprisingly, we find that the impact is limited to the \oidoub\ emission (there is no noticeable effect in the ultraviolet or IR) and amounts to a reduction of order 10\,\% in that line flux. Thus, in the mass range between 11 and about 19\,\msun, this would correspond to an underestimate of the inferred progenitor mass of order 1\,\msun. This error is therefore small. It also suggests that when modeling SNe II at nebular epochs without CO cooling, one ought to select a model that overestimates the \oidoub\ line flux by about 10--20\,\% (an even greater overestimate should be allowed for if CO cooling affects the O/Ne/Mg shell or if SiO also forms). This is what we will do for the modeling of \sn\ in Section~\ref{sect_87A} by selecting model th13.

\subsection{The weak influence of metallicity}
\label{sect_z}

 The new grid of models at one third the solar metallicity returns a paradoxical result in the negligible impact of that metallicity variation on nebular-phase spectra. Figure~\ref{fig_th15_s15p2_350d} illustrates this weak dependence for model th15 (this work) and the solar metallicity counterpart s15.2 \citep{sukhbold_ccsn_16,dessart_sn2p_21} -- this s15.2 model was recomputed with a better model atom and more accurate composition in \citet{dessart_24ggi_25}. The fluxes have been normalized by the decay-power absorbed in each model in order to cancel the slight offset in \nifs\ mass between the  two models (0.063 vs. 0.077\,\msun; see Table~\ref{tab_cmfgen_init}).

 Evidently, the offset between the two models is extremely weak and limited to differences in lines from Fe, Co, and Ni in the IR range, which results from the differences in the yields from explosive nucleosynthesis (e.g., \iso{58}Ni), but also the differences in \ekin\ and \mej. Unlike the strong variations predicted during the photospheric phase \citep{zpap}, and observed (see, e.g., \citealt{anderson_zpap_16}; \citealt{anderson_15bs_18}; \citealt{coulter_lensSN_26}), there is no variation from the forest of Fe lines contributing most of the flux below 5500\,\AA. It will be interesting to explore whether going to one tenth solar would produce an effect. It may be that at nebular times, the spectrum would be indirectly affected by metallicity through changes in preSN evolution, for example with a prolonged and stronger influence of stellar rotation.

\begin{figure}
\centering
\includegraphics[width=\hsize]{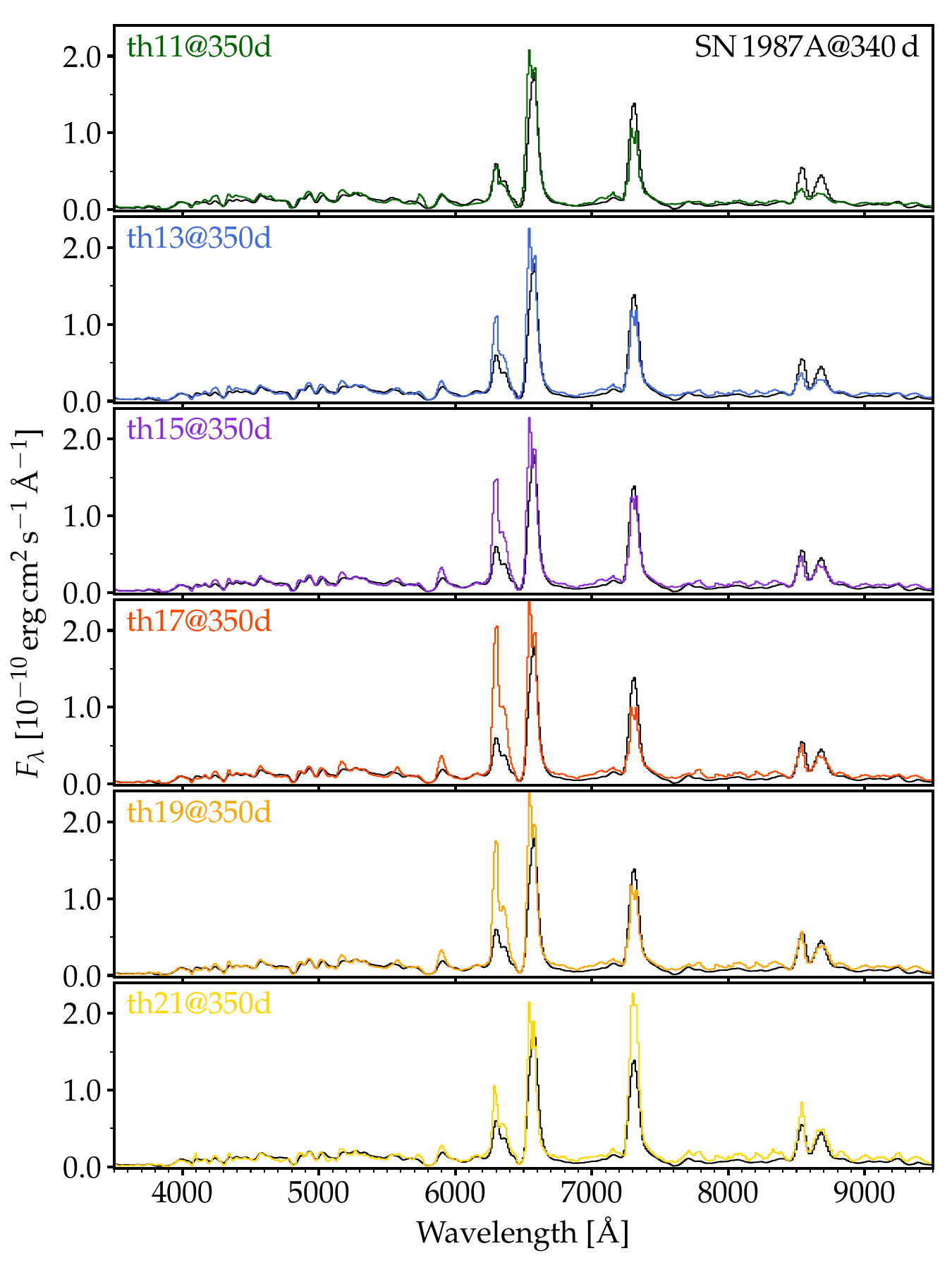}
\caption{Spectral comparison between the observations of \sn\ at 340\,d and models th11 to th21 at 350\,d. The observations have been corrected for reddening and redshift and the models have been scaled to the distance of \sn. No other scaling was applied.
\label{fig_opt_350d}
}
\end{figure}

\section{Comparison to SN\,1987A}
\label{sect_87A}

We now compare the simulations from the above sections to the observations of \sn\ from 200 to 800\,d. We adopted an $E(B-V)$ of 0.15\,mag, a redshift of 0.00096, and a distance of 49.59\,kpc \citep{pietrzynski_lmc_dist_19}. The photometric data that we used for the $V$ band is from \citet{hamuy_87A_90}. We did not include other photometric comparisons because we compare our radiative transfer results to the full spectral energy distribution, covering from UV to the IR whenever possible, and thus our models are compared to data both in absolute and relative terms. The UV spectra from the IUE were taken from \citet{pun_iue_87A_95}.\footnote{We downloaded the IUE spectra of \sn\ from \url{http://ines.oat.ts.astro.it/cgi-ines/IUEdbsMY}.} The optical spectra are from \citet{phillips_87A_90}. The infrared spectra were provided by Jason Spyromilio (priv. comm.) and originate from \citet{meikle_87A_89,meikle_87A_93}, \citet{roche_sio_87A_91}, and \citet{wooden_87A_ir_93}. A slight correction (i.e., at the 10\,\% level) was applied to the optical spectra so that they matched the contemporaneous $V$-band magnitude from \citet{hamuy_87A_90}.

\begin{figure}
\centering
\includegraphics[width=\hsize]{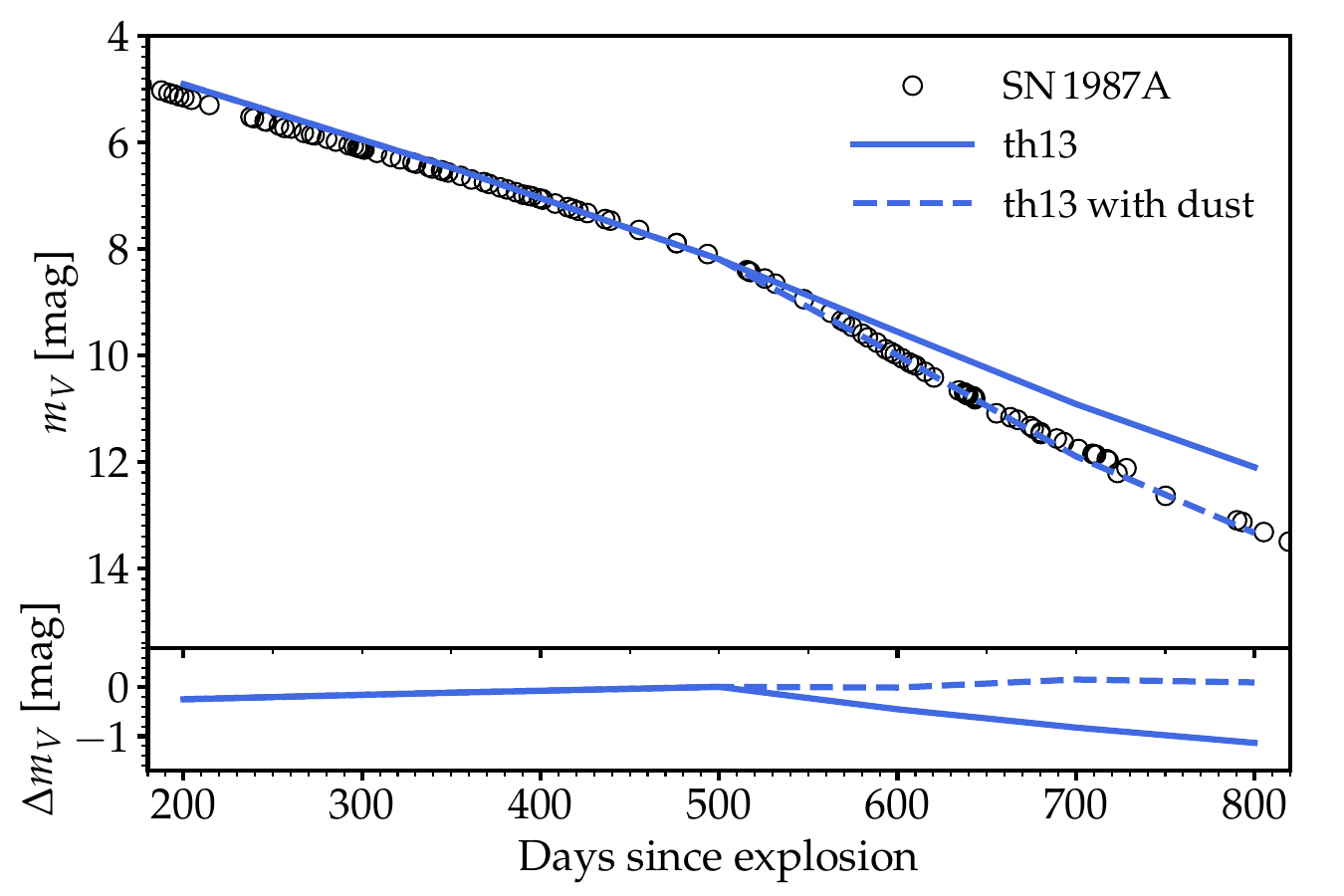}
\caption{Top panel: Comparison of the observed $V$-band light curve of \sn\ with the results for models th13 without (solid) and with (dashed) dust. The model photometry was corrected for extinction and distance. Bottom panel: Same as top but now showing the magnitude offset between observations and models.
\label{fig_Vmag_lc}
}
\end{figure}

Figure~\ref{fig_opt_350d} shows a comparison between the optical spectrum of \sn\ at 340\,d and the model spectra for our set th11 to th21 at 350\,d. Strikingly, all models match the observations throughout the optical with the exception of \oidoub\ in essentially all models. The match is good throughout for th11, but \oidoub\ is increasingly overestimated as we progress from th13 to th17, and that discrepancy is then reduced for the higher initial masses 19 and 21\,\msun. This non-monotonicity was discussed earlier on in Section~\ref{sect_oi}. Since CO cooling is ignored in our simulations, we selected a model that overestimates the \oidoub\ flux by 10-20\,\%, that is model th13. Model th21 could be another option but its large \mej\ may be in tension with the light curve properties.

\begin{figure*}
\centering
\includegraphics[width=0.49\hsize]{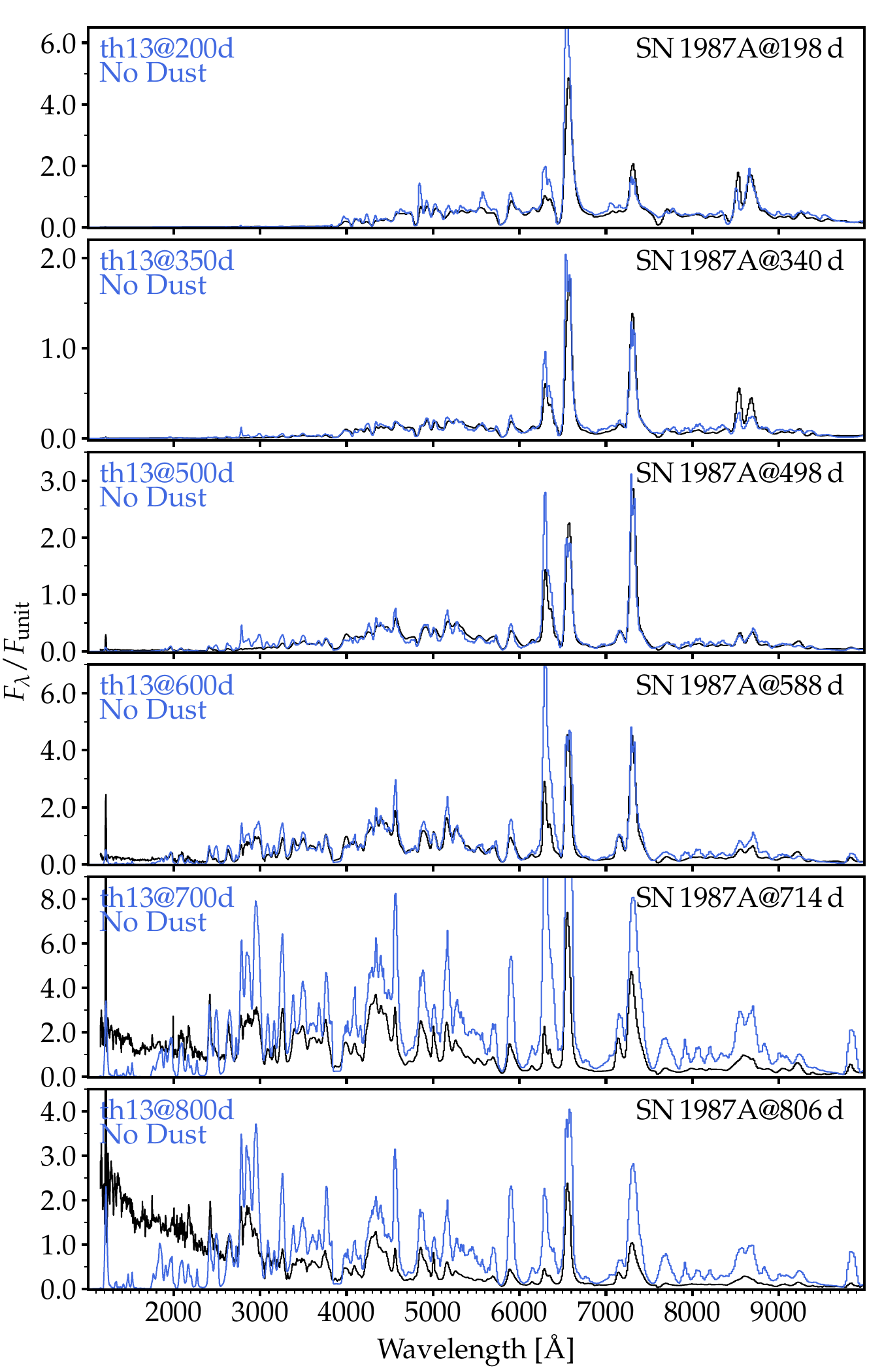}
\includegraphics[width=0.49\hsize]{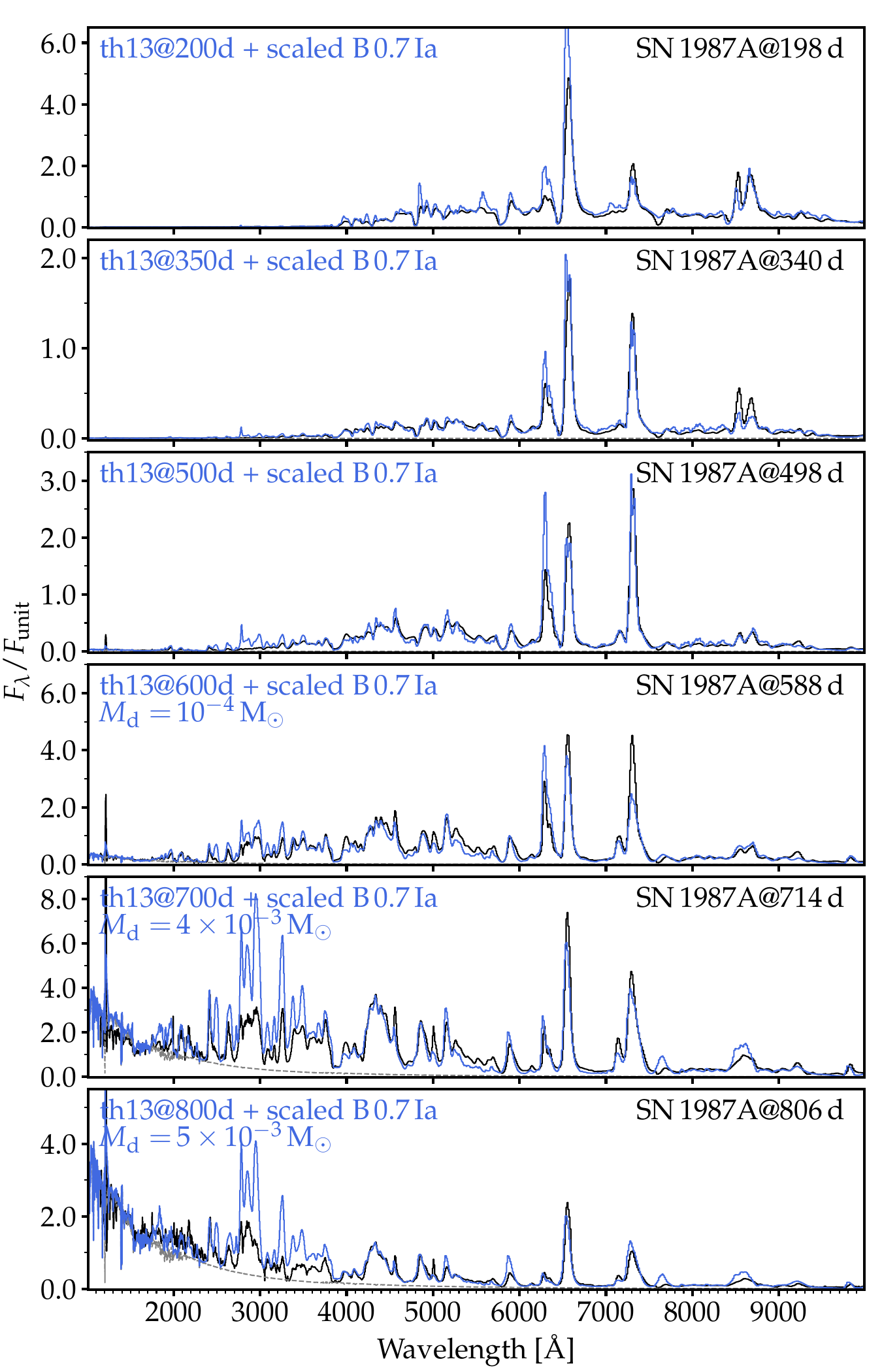}
\caption{Left: Spectral comparison between the observations of \sn\ spanning from 198 until 806\,d and model th13 at roughly contemporaneous epochs -- dust is ignored for the models in that left column. The observations have been corrected for reddening and redshift and the models have been scaled to the distance of \sn. No other scaling was applied. For the ordinate, the quantity $F_{\rm unit}$, in units of erg\,cm$^{-2}$\,s$^{-1}$\,\AA$^{-1}$, is 10$^{-10}$ at 200 and 350\,d, 10$^{-11}$ at 500\,d, 10$^{-12}$ at 600\,d, and 10$^{-13}$ at 700 and 800\,d. Right: Same as left, but first, an increasing amount of dust was added to the model th13 after 500\,d. At all epochs, we added to the SN model spectrum a fixed contamination corresponding to 3\,\% of the flux from a blue supergiant star model (i.e., ``SK-68D140'' from \citealt{alkousa_bsupergiant_lmc_25}). [See Section~\ref{sect_87A} for discussion.]
\label{fig_th13_87A}
}
\end{figure*}

Figure~\ref{fig_Vmag_lc} then shows how this model th13 compares with the $V$-band light curve of \sn. The offset with the data is less than 0.2\,mag at times prior to about 550\,d but then grows to reach 1.2\,mag at 800\,d. The knee in the light curve at 500\,d is not associated with $\gamma$-ray escape since this varies smoothly in time (see Fig.~\ref{fig_edecay}). The left panel of Fig.~\ref{fig_th13_87A} shows the spectral evolution of this same model th13 from 200 to 800\,d. The model spectra are in agreement with observations until 500\,d and subsequently deviate, with eventually a strong discrepancy at 800\,d. The main offset is not a poor match of line profile shapes. It is instead primarily a strong overestimate of the observed flux by a factor two or more throughout the optical range.

As indicated by \citet{lucy_dust_89}, newly-formed dust in the inner, metal-rich ejecta of \sn\ is at the origin of this unexpected change in radiative properties of \sn. Since small silicate dust grains are likely to form first at about 500\,d in Type II SNe \citep{sarangi_dust_22}, we recomputed the model th13 after distributing 0.1\,\mic\ silicate dust grains uniformly below 2000\,\kms. For each epoch from 500 to 800\,d, we computed cases with a dust mass $M_{\rm d}$ between 10$^{-4}$ and $5 \times 10^{-3}$\,\msun. In Fig.~\ref{fig_Vmag_lc} and the right column of Fig.~\ref{fig_th13_87A}, we compare these dusty th13 models with the $V$-band light curve and the UV and optical spectra of \sn. By invoking a dust mass of 10$^{-4}$, $4 \times 10^{-3}$, and $5 \times 10^{-3}$\,\msun, at 600, 700, and 800\,d, the now dusty model th13 matches well the $V$-band light curve with an offset from observations that is within 0.2\,mag at all times. We also obtain a satisfactory match of the spectra throughout the optical range, in particular with the strong reduction in \oidoub\ and the presence of a relatively strong Fe\two\ emission below 5500\,\AA. Evidently, the small blue-red profile asymmetries introduced by dust are not perfectly fitted (for a detailed modeling of the impact of dust on \ha, see, for example, \citealt{bevan_barlow_16}) but our model yields a global consistency -- the global reduction of the flux by dust is also much greater than the blue-red asymmetry on line profiles (which may also encode potential asymmetries of the ejecta density and composition). We also tested with an increased outer velocity bounding the dust at 3000\,\kms\ but this caused a more global attenuation of the spectra (the entire ejecta emission suffered dust attenuation) whereas the observations support a differential influence of the dust, with a greater impact on the emission from the metal-rich ejecta and less so of the H-rich material.

Not discussed so far is the rising UV flux (i.e., relative to the optical) in the IUE data of \sn\ at late nebular times \citep{pun_iue_87A_95}. The UV flux is vanishingly small prior to 500\,d apart from a weak \lya\ emission. At 600\,d, a flat UV flux is clearly present, and it continues to grow until the epoch shown here at 800\,d. While the UV flux increases in the (dust-free) model th13 at those late times (Fig.~\ref{fig_cum}), it does not rise by as much. To match the observed evolution in the IUE range, we introduced a small fraction of the flux (i.e., 3\,\%) from the blue supergiant spectrum model ``SK-68D140'' of \citet{alkousa_bsupergiant_lmc_25} characterized by an effective temperature of 24\,100\,K, a $\log g$ of 2.81, and a $\log L/L_\odot$ of 5.52. \citet{pun_iue_87A_95} suggested that the IUE spectra were contaminated by background stars after $\sim$\,900\,d but our radiative transfer calculations for \sn\ suggest that this contamination is clearly present at $\gtrsim$\,600\,d -- inspecting the spectra in log-space indicates that this contamination is present 200\,d onwards (see below, the appendix, and in particular Fig.~\ref{fig_87A_th13_200d_350d}--\ref{fig_87A_th13_700d_800d}).

This contamination is not surprising. At late times, the bolometric luminosity drops to $\sim$\,10$^{40}$\,\ergs\ at 500\,d and is as low as $\sim$\,10$^{38}$\,\ergs\ at 800\,d. The latter is comparable to the luminosity of any supergiant star and it is clear that the SN luminosity may be drowned by contaminating light. However, this contamination appears much earlier in the UV, when the SN is more luminous but UV faint, in particular if the contamination arises from a hot star like a blue supergiant. A hot (i.e., of order 20\,kK) massive star can then outshine \sn\ in the UV at nebular times. This contamination would be weak in the UV and stronger in the optical if the contaminating star was a red supergiant or a cluster of such cooler stars.

\begin{figure}
\centering
\includegraphics[width=\hsize]{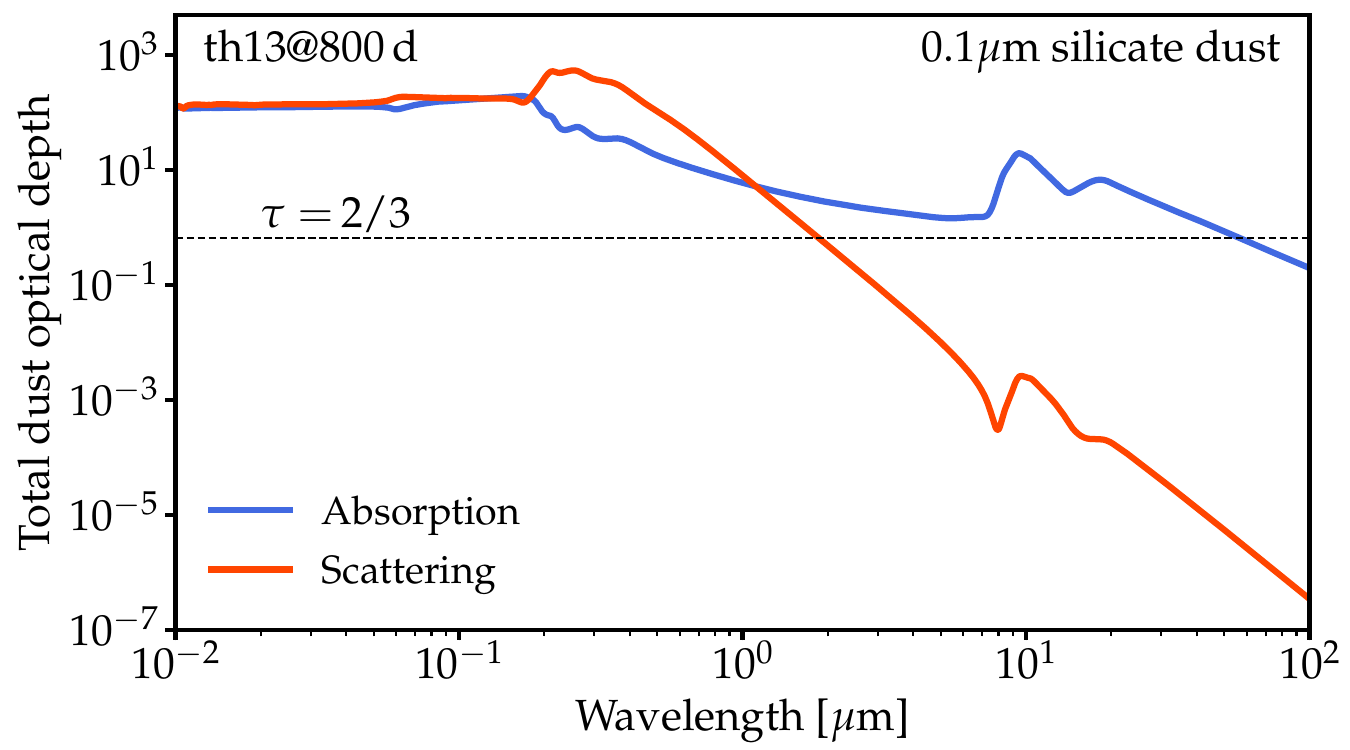}
\caption{Variation of the total ejecta dust optical depth associated with scattering and absorption in model th13 at 800\,d for $5 \times 10^{-3}$\,\msun\ of 0.1\,\mic\ silicates distributed uniformly in the inner ejecta below 2000\,\kms.
\label{fig_tau_dust}
}
\end{figure}

In the appendix, we show additional illustrations of the model comparison to the UV, optical, and IR observations of \sn, together with variation of the dust temperatures at 200, 250, 300, 350, and 400\,K (Figs.~\ref{fig_87A_th13_200d_350d}--\ref{fig_87A_th13_700d_800d}). Higher dust temperatures are necessary at earlier times (i.e., $\sim$\,400\,K at 600\,d compared to 250\,K at 700 and 800\,d), although there might be a SiO contribution to the IR emission, thus unrelated to thermal dust emission  \citep{roche_sio_87A_91,wooden_87A_ir_93}. In our simulations of dust in the model th13, we find that the dust is optically thick throughout the spectrum from the ultraviolet to the infrared. Figure~\ref{fig_tau_dust} illustrates the evolution of the total ejecta dust optical depth in model th13 at 800\,d for $5 \times 10^{-3}$\,\msun\ of 0.1\,\mic\ silicates distributed uniformly in the inner ejecta below 2000\,\kms. This leads to a more modest IR emission compared to optically-thin conditions for the same dust mass, but also means that blue-red profile asymmetries as well as flux attenuation affects all lines in the UV and IR. As shown in Figs.~\ref{fig_87A_th13_200d_350d}--\ref{fig_87A_th13_700d_800d}, with our choice of dust, we tend to overestimate the blue-red asymmetry of NIR line profiles. Reducing the dust mass may improve the match in the NIR but would deteriorate the match in the optical at 600--800\,d. An alternative could be to reduce the grain size to 0.01\,\mic, as this would boost the opacity in the UV relative to the optical and IR (our models do tend to overpredict the flux around 3000\,\AA), but this would introduce a mismatch in the optical where the dust opacity would then be smaller. Given all these complications and the relatively good agreement of our models with the observations of \sn, we leave it at that. The comparison of the dusty-model th13 with \sn\ also shows a rough agreement to the main lines present in the IR (Figs.~\ref{fig_87A_th13_200d_350d}--\ref{fig_87A_th13_700d_800d}), but it is difficult to be more specific given the very low resolution and poor signal of the observations in that wavelength range.

\section{Conclusions}
\label{sect_conc}

  We have presented nonlocal thermodynamic equilibrium time-dependent radiative-transfer models for ejecta resulting from the explosion of progenitor stars initially between 11 and 21\,\msun\ and evolved at $Z_\odot/3$. Since the original aim was to confront to observations of \sn, we adopted an \ekin\ of $1.2 \times 10^{51}$\,erg and a \nifs\ mass of 0.07--0.08\,\msun. Macroscopic mixing in the ejecta was enforced by adopting a shuffled-shell composition structure. The times considered covered from 200 to 800\,d. Five two-step decay chains were included, including those led by \nifs\ and \iso{57}Ni. Molecular formation and cooling were ignored but dust in the form of 0.1\,\mic\ silicate grains was introduced as a post-treatment stage. In contrast to previous similar work, the possibility of interaction with circumstellar material was ignored in order to focus on the evolution of the emission from the inner, metal-rich and decay-powered ejecta -- no other power source was included. Previous work like \citet{dessart_ir_25} studied similar configurations but were limited to earlier times (i.e., $\lesssim$\,500\,d), to solar-metallicity progenitors, ignored dust, and did not include any detailed comparison to observations. However, we found little difference in results with the solar-metallicity counterparts so that aspect is little studied in this work -- going to even lower metallicity ($Z_\odot/10$?) or working with rotating stars may lead to visible variations but this is left to future work.

  Our simulations follow a considerable drop in bolometric luminosity from $\sim$\,10$^{41}$\,\ergs\ at 200\,d down to $\sim$\,10$^{38}$\,\ergs\ at 800\,d, and thus down to luminosities typical not of SNe but of standard massive stars. Over that time span, $\gamma$-ray escape increases from essentially zero to 50--80\,\%, the more so for lower \minit\ (all models have the same \ekin\ but differ in \mej), with a growing fraction of positrons from 3\,\% to 6--13\,\%, and an increasing (decreasing) fraction of the decay power absorbed by the H-rich (O-rich) material. While the optical range contains the bulk of the flux at all times, both the UV and the IR contain a growing fraction of the total. In some sense, the spectral energy distribution increasingly spreads out in wavelength from UV to IR as time passes.

  The \oidoub\ line flux shows a complicated trend in our simulations. It shows a rise from 200 to 500--600\,d followed by a decline, probably as a result of the decrease of the decay power absorbed in the O-rich material. The fractional line flux in \oidoub\ grows for increasing \minit\ up to 17--19\,\msun, but it decreases for larger masses as a result of the greater density in the inner ejecta of more massive progenitors. This non-monotonicity compromises the use of that line flux for progenitor mass determination, in particular in its ability to exclude higher-mass progenitors as the source of SNe II. These results reveal an important degeneracy of nebular-phase spectra of SNe II.

  We also evaluated the potential impact that CO cooling would have on nebular-phase spectra. Using the simplistic assumption that CO molecules would cause 90\,\% of the cooling of the C/O shell, we found that the impact was essentially limited to the \oidoub\ line flux, with a small reduction of 10--20\,\%. This amounts to a potential underestimate of the inferred \minit\ by about 1\,\msun\ for models below 17\,\msun\ and that ignore CO cooling. CO cooling may also affect the O/Ne/Mg shell and would in that case quench further the [O\one] line flux as well as IR lines like \neiifs. And such effects would be exacerbated with the formation of SiO and the associated cooling. All such coolants would tend to reduce the [O\one] line flux and lead to an underestimate of \minit\ -- this would allow a higher mass model (i.e., th15 or perhaps th17) as a better match for \sn\ than our current better-match model th13.

  The comparison to \sn\ led to some surprises. As has been advocated by single-line studies, the blue-red asymmetry of line profiles like \ha\ after 500\,d requires the presence of dust. In our best match model th13, we computed dust radiative transfer models from UV to IR and found that dust is indeed essential to match not just profile asymmetries but the global attenuation of the entire spectrum, by about 1.5\,mag in the $V$-band. The observed fading in \sn\ at 500\,d cannot arise from $\gamma$-ray escape, which has a more gradual evolution. Invoking an increase in mass of 0.1\,\mic\ silicates from about 10$^{-4}$\,\msun\ at 600\,d to a few 10$^{-3}$\,\msun\ at 800\,d reconciles the light curve and spectra of model th13 with \sn. The greater attenuation of lines like \oidoub\ compared to \ha\ or Fe emission from the H-rich material requires the dust to be relatively deep in the ejecta, typically below 2000\,\kms\ -- placing it below 3000\,\kms\ deteriorates the model agreement with \sn. Some discrepancies remain for some lines, for example around 3000\,\AA, which arises from emission by Fe lines at the base of the outer H-rich ejecta.

  Given all these uncertainties, we thus adopted our (dust-forming) 13\,\msun\ $Z_\odot/3$ model as best matching the nebular phase properties of \sn. Its O yield is 0.79\,\msun\ and thus smaller than inferred by early studies of \sn\ \citep{spyromilio_pinto_91,li_mccray_o1_92,chugai_87A_94}, but the level of physical consistency of the present work is also greater. Furthermore, we neglect any clumping of the ejecta material, which might also seem to contradict those studies. In reality, this arises in part from a conflict in language. Indeed, even though we neglect clumping, that is we do not invoke any process that could compress any material (e.g., we neglect the \nifs\ bubble effect), there is strong segregation of material in the adopted shuffled-shell composition of our models. For example, in model th13 the volume bounded by a velocity of 2500\,\kms\ is occupied at a 37\,\% level by H-rich material, at a 29\,\% level by He-rich material, at a 30\,\% level by O-rich material, and at a 4\,\% level by Fe-rich material. That is, the volume filling factor of O-rich material is 30\,\%, and thus not so dissimilar to the 10\,\% volume filling factor inferred by \citet{spyromilio_pinto_91}. However, this is not clumping since the term generally implies some compression of the gas. This configuration instead reflects the inhomogeneous nature of the composition within the inner, metal-rich regions of the ejecta (Fig.~\ref{fig_prog}). In fact, the original progenitor composition at core collapse is similarly characterized by distinct shells (i.e., H-rich, He-rich etc) in mass space so that only a fraction is occupied by each shell within some given Lagrangian mass. Whether this material is organized in spherical shells as in our 1D setup or in individual, uncompressed blobs in the ejecta is simply a question of spatial arrangement.

  A 13\,\msun\ progenitor lies at the lower end of the proposed range in the literature. The progenitor luminosity came chiefly from the helium shell and supported the He core mass to be 5 to 7\,\msun\ from the inferred luminosity of $3-6 \times 10^{38}$\,\ergs\   \citep{sn1987A_rev_90}. The model th13 has a He-core mass of 4.04\,\msun, and a luminosity of $3.21 \times 10^{38}$\,\ergs\ at core collapse, so low but within the bounds \citep{woosley_87A_late_88}. We conducted tests with clumping (i.e., compression) of the O-rich material and found that it can contribute to reducing the \oidoub\ line flux, but at the expense of \naid\ and \mgifs, which would then be overpredicted relative to \sn. Revisions upward of this progenitor mass might arise from a proper treatment of molecular cooling, since both CO and SiO would drain some of the \oidoub\ flux. Fully coupled molecule formation and cooling within the radiative transfer is however challenging in Type II SN ejecta \citep{mcleod_co_24}.

  Importantly, we found that the IUE spectra of \sn\ are dominated by a contamination from a background hot star at all times in the 200--800\,d interval considered here. Indeed, although our models brighten in the UV relative to the optical as time passes, the observed UV brightening (in a relative sense) of \sn\ is much greater. By adding a constant flux from a B\,0.7\,Ia supergiant model (scaled by 0.04), we reconcile models and UV observations of \sn. Such contamination is typically ignored in the interpretation of observations (although it is clearly a recognized challenge) but could be a major limitation for the understanding of late time observations of Type II SNe at larger distances than \sn. Once or if these have reached down to powers of 10$^{38}$\,\ergs, the potential contamination in the UV from nearby blue supergiants or in the optical from red supergiants becomes a concern.

\begin{acknowledgements}

LD thanks Stan Woosley for fruitful discussions and for computing the stellar evolution and explosion models used as initial conditions for the radiative transfer calculations presented in this work.
LD thanks Jason Spyromilio for providing the IR data of \sn.
LD acknowledges support from the ESO Scientific Visitor Program for a visit to ESO-Garching during the summer of 2025 and during which this work was initiated.
This work was granted access to the HPC resources of TGCC under the allocations 2024 -- A0170410554 and 2025 -- A0190416871 on Irene-Rome made by GENCI, France.
This research has made use of NASA's Astrophysics Data System Bibliographic Services.

\end{acknowledgements}

\bibliographystyle{aa}
\bibliography{new_sn_library_luc}

\appendix

\onecolumn

\section{Additional illustrations}
\label{sect_appendix}

\begin{figure*}[b!]
\centering
\includegraphics[width=0.7\hsize]{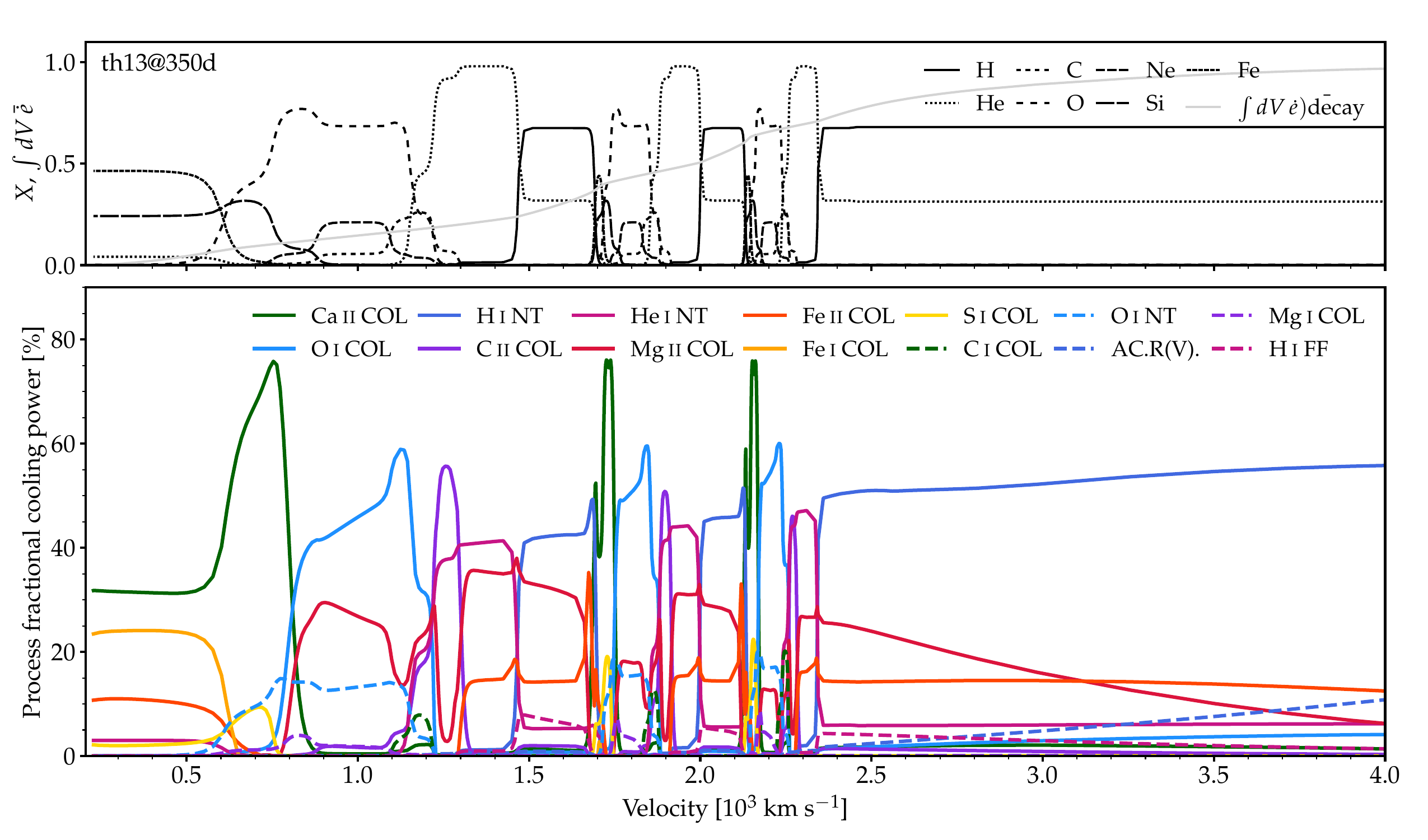}
\includegraphics[width=0.7\hsize]{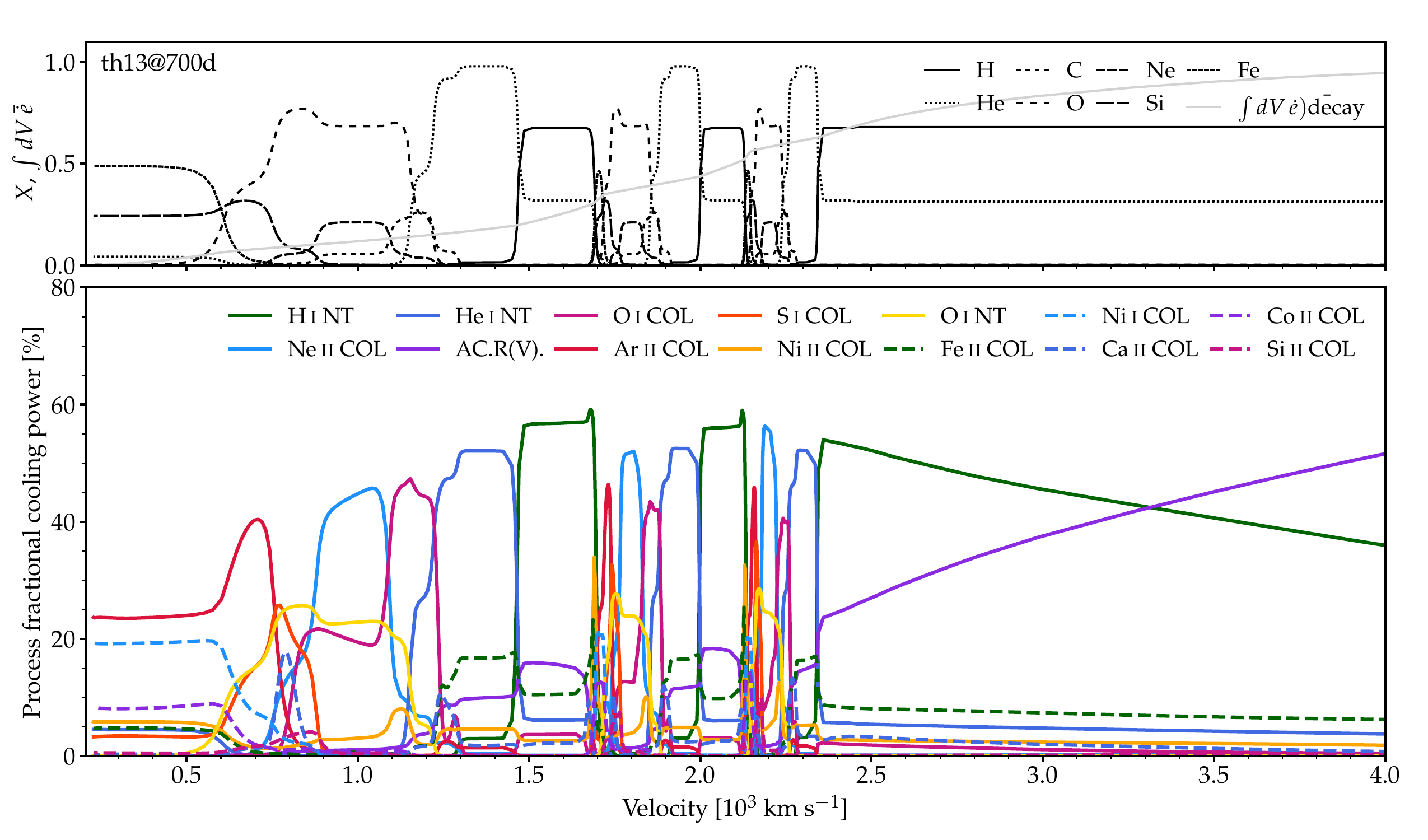}
\caption{Top: Illustration of ejecta and radiative properties for model th13 at 350\,d. The top row shows the mass fraction of H, He, C, O, Ne, Si, and Fe as a function of velocity, which reveals the location and extend of the main shells. Also shown is the normalized cumulative absorbed decay power integrated outwards from the innermost ejecta layer (light grey line). The bottom row shows the variation of the local cooling power, as a percentage of the local total heating power, for the top 14 coolants ejecta-wide. The term ``COL'' refers to collisional excitation, ``FF'' to free-free emission,  and ``NT'' to nonthermal excitation. The term ``AC.R(V).'' refers to cooling from expansion. Bottom: Same as top but now at 700\,d.
\label{fig_cool_th13_350d_700d}
}
\end{figure*}

\begin{figure*}
\centering
\includegraphics[width=\hsize]{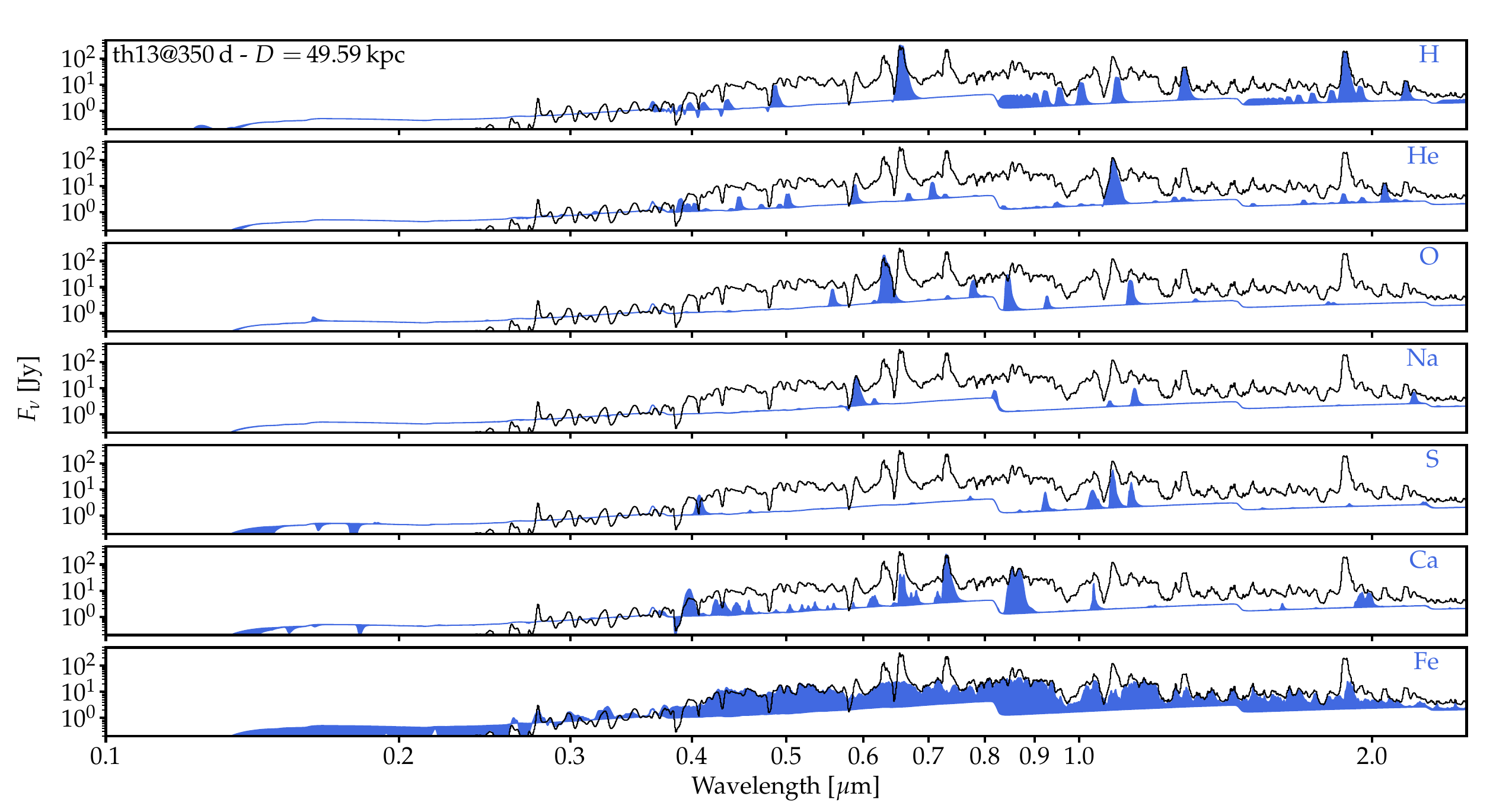}
\caption{Spectral decomposition of model th13 at 350\,d showing the main contributors to the spectrum. The spectra were scaled to the distance of \sn. The shading corresponds to the flux arising from the corresponding species (i.e., all ions from the species indicated at top right of each panel) relative to the continuum flux. Over wide spectral regions, this species-dependent flux deviates from the local flux because of overlap with other species (e.g., absorption or emission from Fe lines etc).
\label{fig_th13_350d_spec}
}
\end{figure*}

\begin{figure*}
\centering
\includegraphics[width=\hsize]{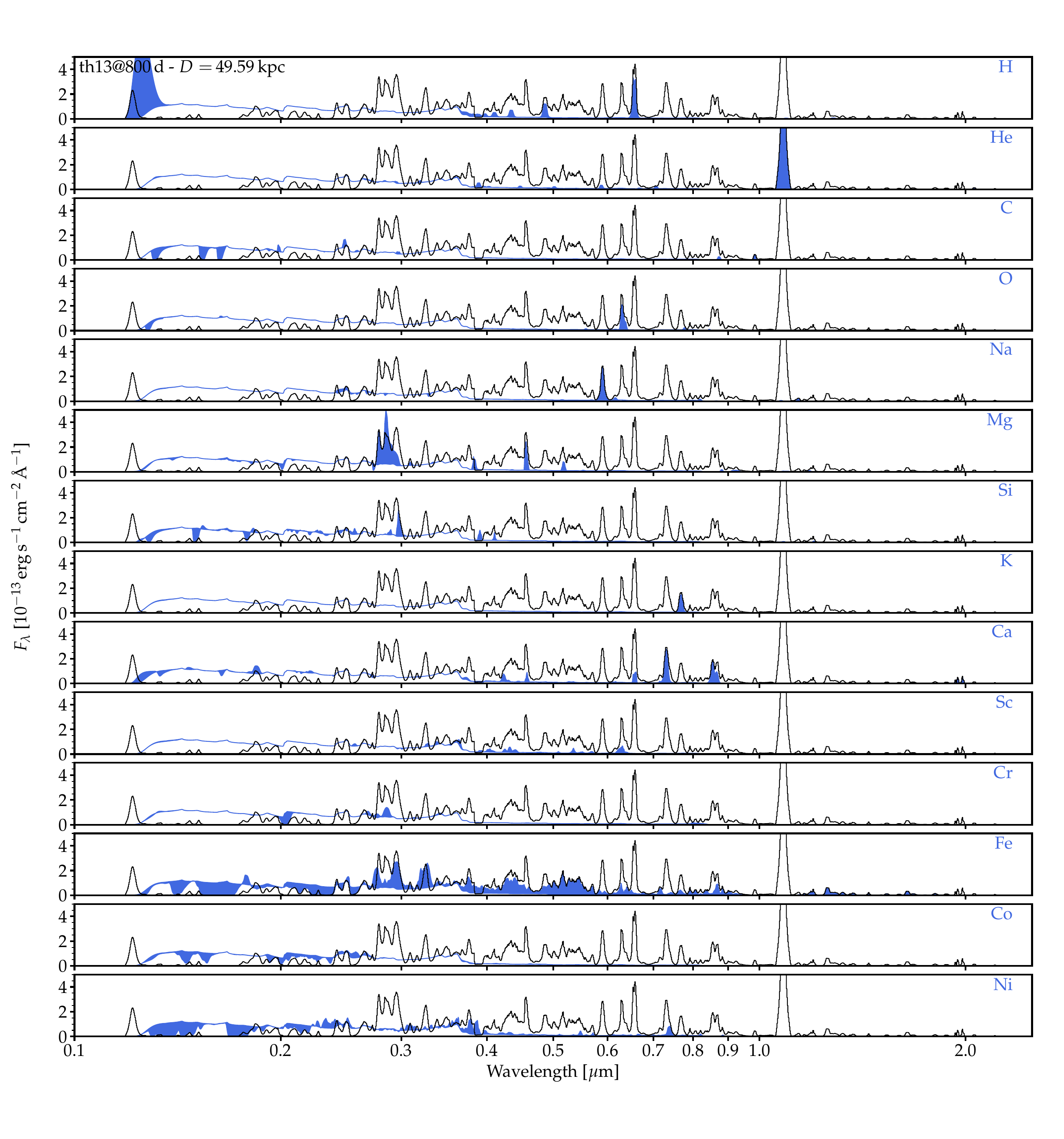}
\caption{Same as Fig.~\ref{fig_th13_350d_spec} but now for model th13 at 800\,d.
\label{fig_th13_800d_spec}
}
\end{figure*}

\begin{figure*}
\centering
\includegraphics[width=\hsize]{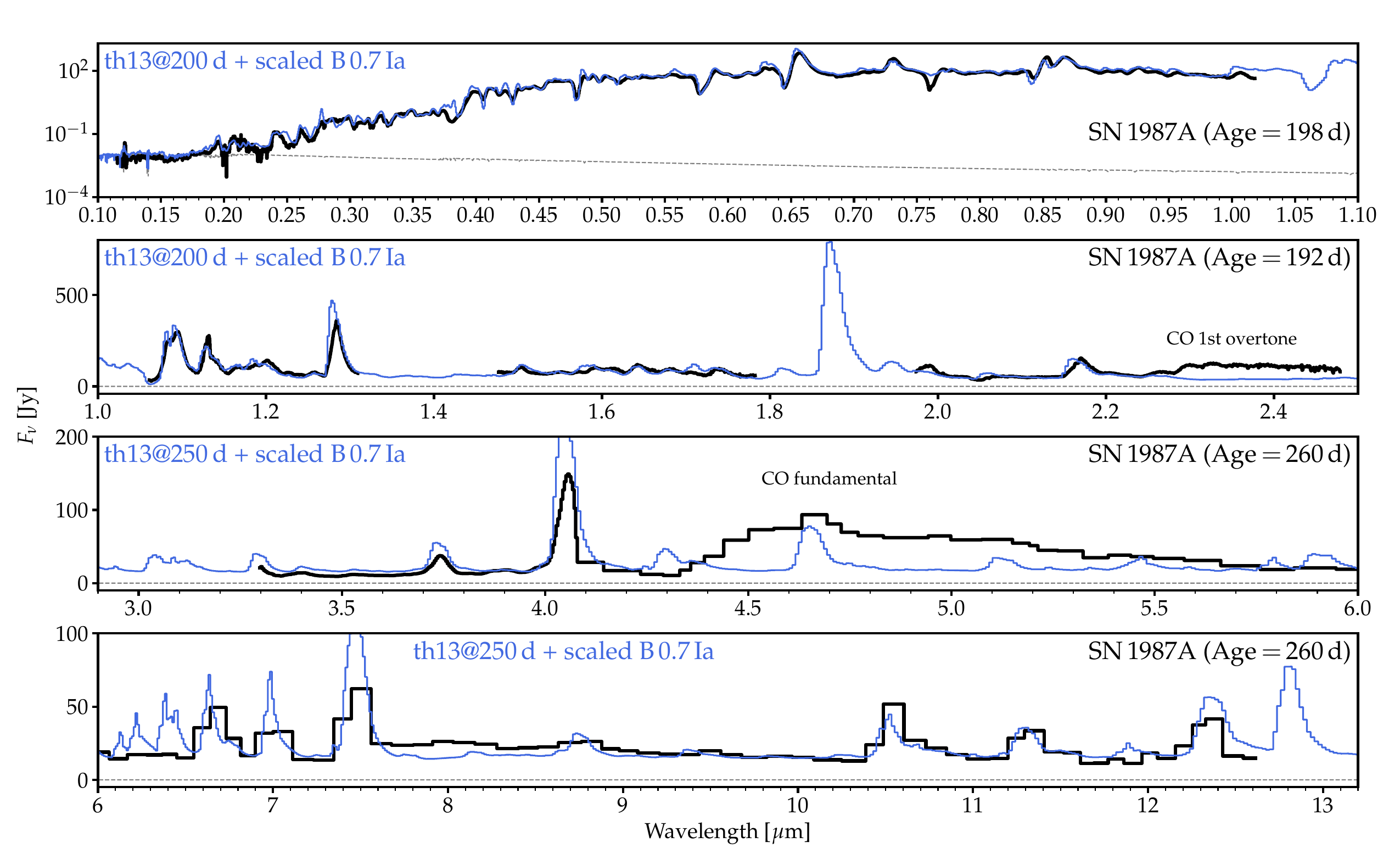}
\includegraphics[width=\hsize]{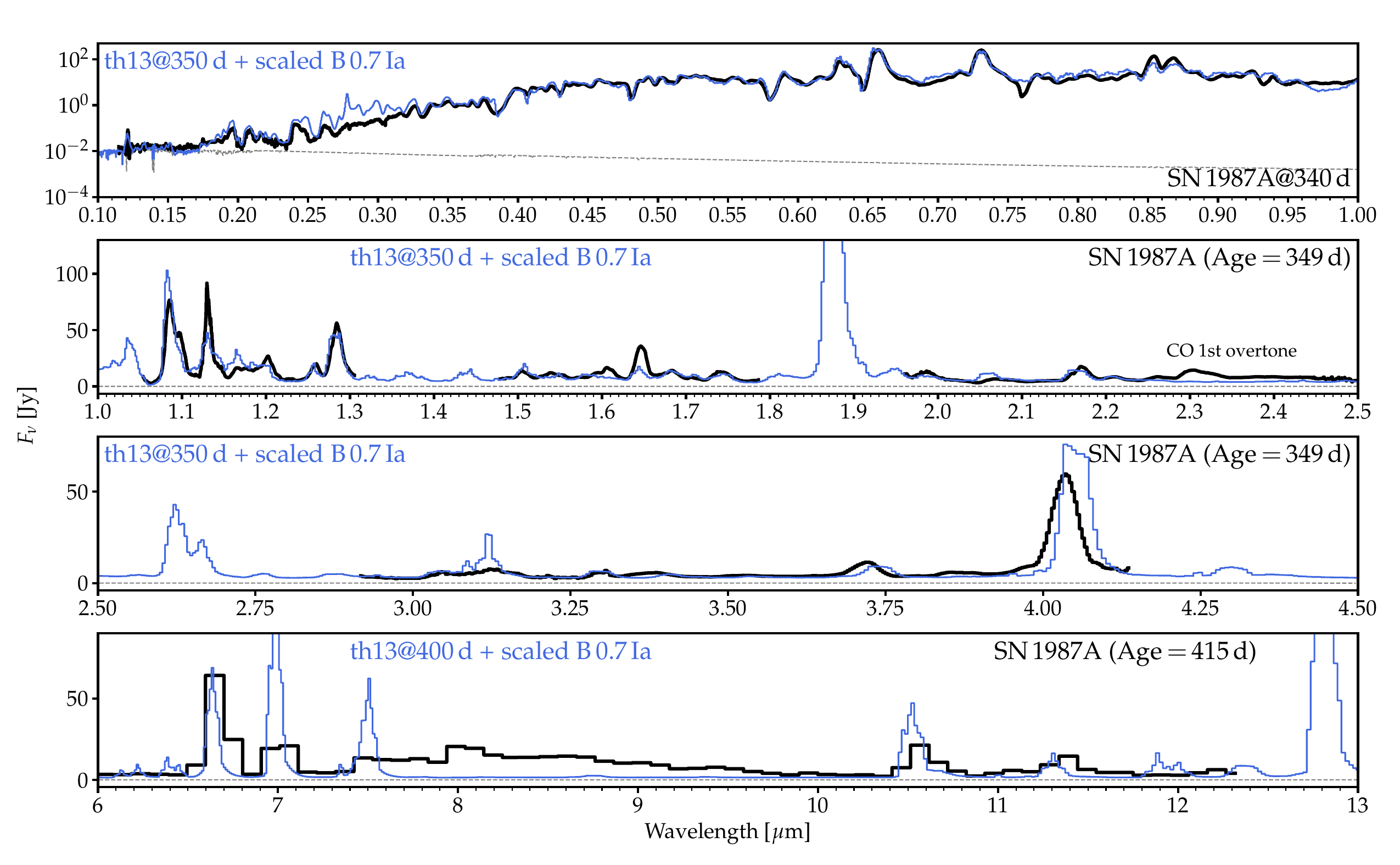}
\caption{Spectral comparison between \sn\ and model th13 at about 200\,d (top) and 350\,d (bottom). The observations were corrected for redshift and reddening and the models were scaled to the distance to \sn\ as well as rebinned to a resolution of 500\,\kms. Dust is ignored at both epochs but the model spectra are augmented with a fixed (non-time-evolving)  contribution amounting to 3\,\% of the flux from the B\,0p7\,Ia model named ``SK-68D140'' \citep[shown as a dashed grey line]{alkousa_bsupergiant_lmc_25}.
\label{fig_87A_th13_200d_350d}
}
\end{figure*}

\begin{figure*}
\centering
\includegraphics[width=\hsize]{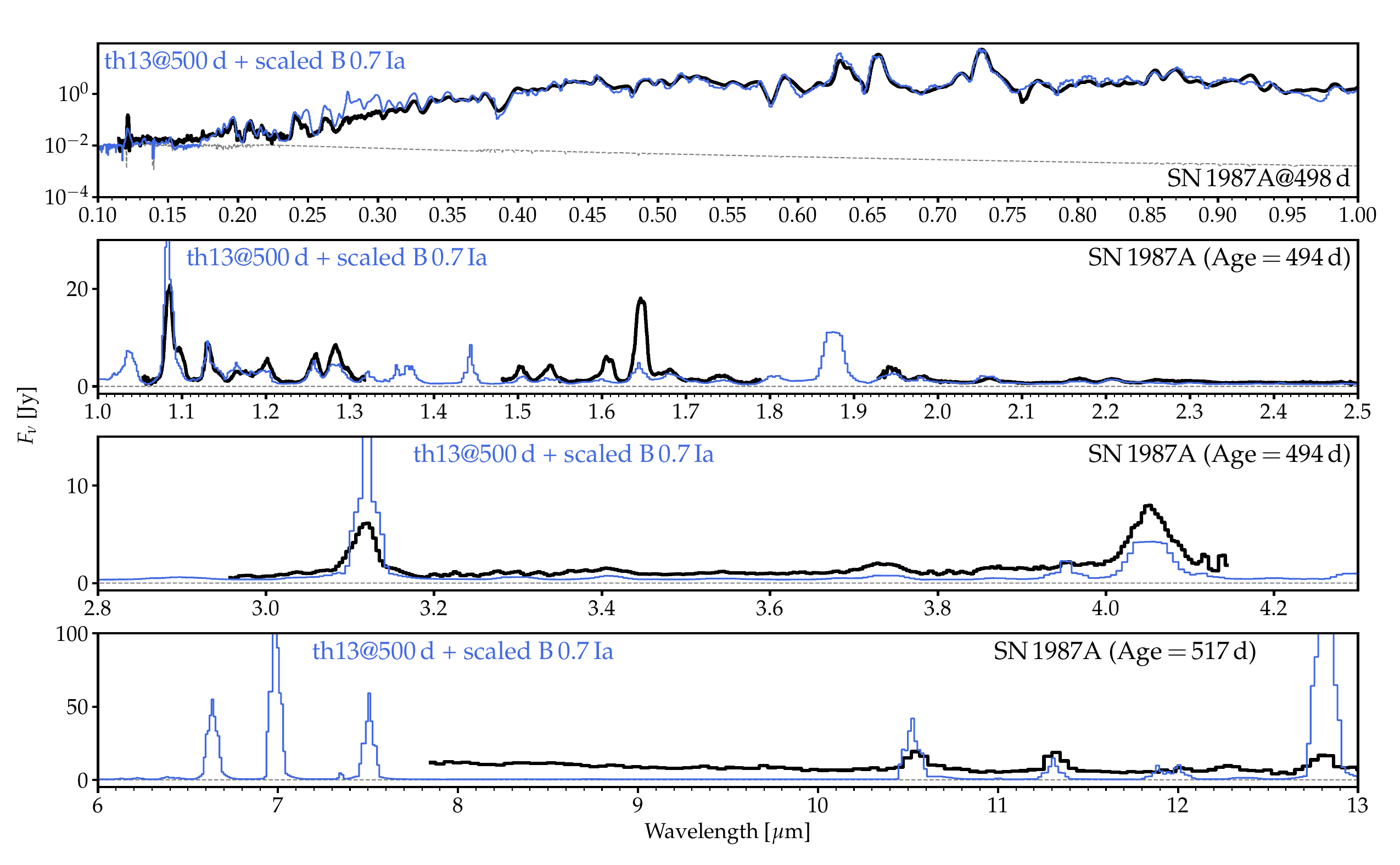}
\includegraphics[width=\hsize]{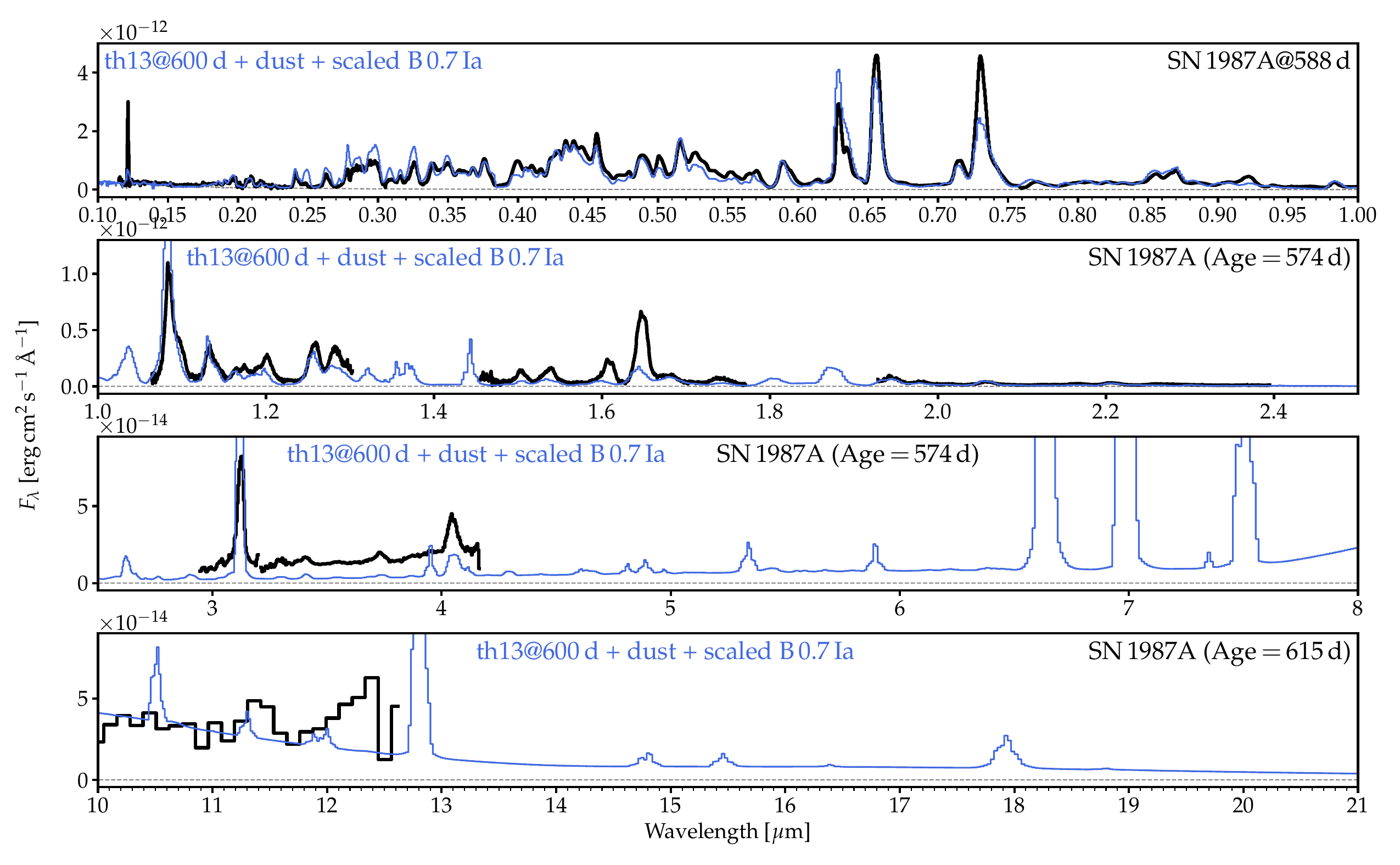}
\caption{Same as Fig.~\ref{fig_87A_th13_200d_350d} but now at 500\,d (top) and 600\,d (bottom). The 600\,d model includes dust from 10$^{-4}$\,\msun\ 0.1\,\mic\ silicates at 400\,K and distributed uniformly below 2000\,\kms.
\label{fig_87A_th13_500d_600d}
}
\end{figure*}

\begin{figure*}
\centering
\includegraphics[width=\hsize]{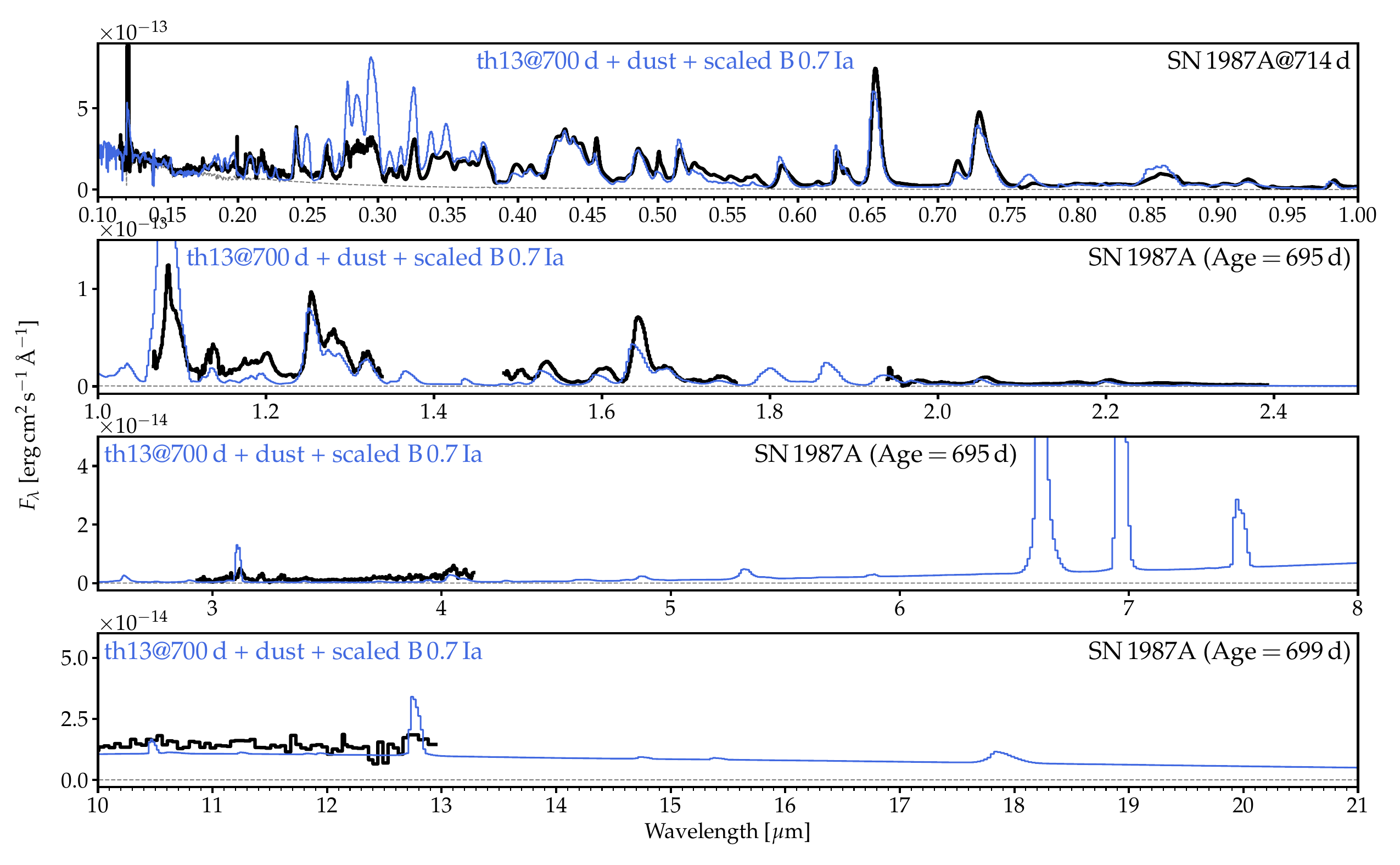}
\includegraphics[width=\hsize]{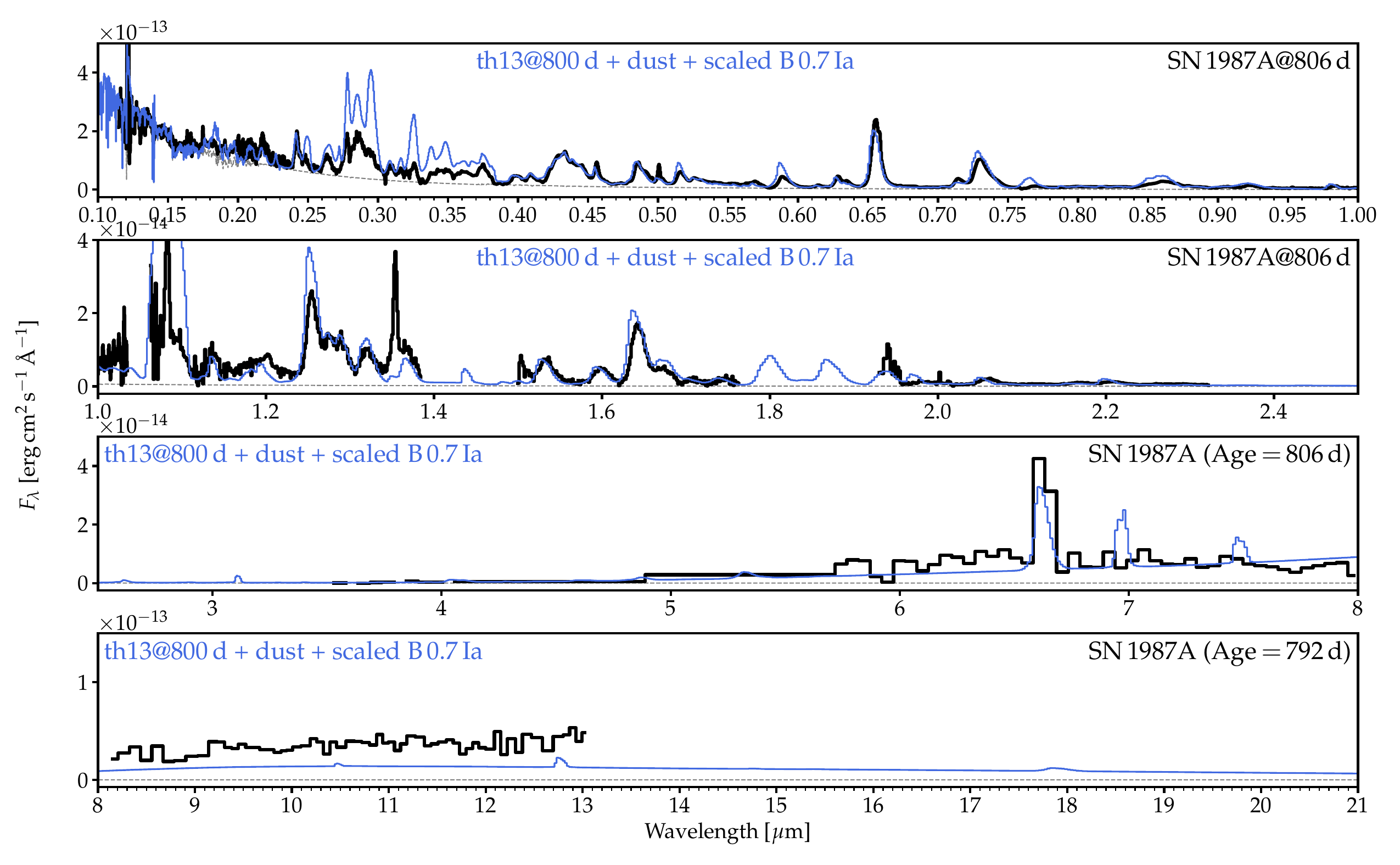}
\caption{Same as Fig.~\ref{fig_87A_th13_200d_350d} but now at 700\,d (top) and 800\,d (bottom). Both models include dust from 4 and $5 \times 10^{-3}$\,\msun\ 0.1\,\mic\ silicates at 250\,K and distributed uniformly below 2000\,\kms.
\label{fig_87A_th13_700d_800d}
}
\end{figure*}

In this appendix, we provide additional information to complement what was discussed in the main text. Figure~\ref{fig_cool_th13_350d_700d} present the main coolants in different regions of the ejecta, primarily controlled by the large variations in composition in the adopted shuffled-shell structure of our models. Figures~\ref{fig_th13_350d_spec} and \ref{fig_th13_800d_spec} present a break down of the main contributors to the (dust-free) th13 spectrum from UV to IR at 350 and 800\,d. While elements of low atomic weight tend to contribute only a few lines, heavier elements and in particular those from the iron group contribute forests of lines. Figures~\ref{fig_87A_th13_200d_350d}--\ref{fig_87A_th13_700d_800d} present a UV-to-IR comparison of the best-matching th13 model, augmented by a contaminating blue supergiant star model and at times by dust, and at post-explosion epochs of 200, 350, 500, 600, 700, and 800\,d. For better visibilty, the spectra may be shown in the log, as $F_\lambda$, or as $F_\nu$.

\end{document}